\documentclass[twocolumn,nofootinbib,amsmath,prd,aps,superscriptaddress,tightenlines,preprintnumbers]{revtex4}

\usepackage{graphicx}
\usepackage{braket}
\usepackage[usenames]{color}
\usepackage{latexsym}
\usepackage{amsmath,accents}
\usepackage{mathrsfs}
\usepackage{hyperref}
\usepackage{enumerate}
\usepackage{adjustbox}
\usepackage{amssymb}
\usepackage{cancel}

\def\bea{\begin{eqnarray}}
\def\eea{\end{eqnarray}}

\preprint{FTUV-17-1110.0001}
\preprint{IFIC-17/44}

\begin{document}
\newcount\hour \newcount\minute
\hour=\time \divide \hour by 60
\minute=\time
\count99=\hour \multiply \count99 by -60 \advance \minute by \count99
\newcommand{\mydate}{\ \today \ - \number\hour :00}

\title{LHC signals of radiatively-induced neutrino masses \\and implications for the Zee-Babu model}

\author{Julien Alcaide, Mikael Chala and Arcadi Santamaria\\\vspace{0.4cm}
\it {Departament de F\'isica T\`eorica, Universitat de Val\`encia and IFIC, Universitat de
Val\`encia-CSIC, Dr. Moliner 50, E-46100 Burjassot (Val\`encia), Spain}
}

\begin{abstract}
   Contrary to the see-saw models, extended Higgs sectors leading to radiatively-induced neutrino masses do require the extra particles to be at the TeV scale. However, these new states have often exotic decays, to which experimental LHC searches performed so far, focused on scalars decaying into pairs of same-sign leptons, are not sensitive. In this paper we show that their experimental signatures can start to be tested with current LHC data if dedicated multi-region analyses correlating different observables are used. We also provide high-accuracy estimations of the complicated Standard Model backgrounds involved. For the case of the Zee-Babu model, we show that regions not yet constrained by neutrino data and low-energy experiments can be already probed, while most of the parameter space could be excluded at the 95 \% C.L. in a high-luminosity phase of the LHC.
\end{abstract}

\maketitle
\newpage
\tableofcontents

\section{Introduction}

The Standard Model (SM) gives a natural and simple explanation for exactly massless neutrinos. However, neutrino oscillation data provide an irrefutable evidence of neutrino masses, which are much smaller than the rest of the fermions. Neutrino masses can be accommodated in extensions of the SM involving new particles or parametrized by non-renormalizable operators violating lepton number conservation (LN). 

The most straightforward extension is obtained by adding three families of singlet right-handed neutrinos. If they have a large Majorana mass term $M$, LN is broken, left-handed neutrino masses are generated at tree level and their smallness can naturally be explained by the see-saw formula  $m_{\nu}\sim v^2/M$, with $v$ the vacuum expectation value (VEV) of the SM Higgs doublet. This is the so-called see-saw model type I~\cite{Minkowski:1977sc,GellMann:1980vs,Yanagida:1979as,Mohapatra:1979ia} which arises in Grand Unification models like the ones based on $SO(10)$. The simplicity of the model and the fact that it appears naturally in well motivated extensions of the SM makes the see-saw type I as the most appealing explanation of neutrino masses. A more complicated extension, but richer phenomenologically, is obtained by replacing the singlet neutrinos by triplets of fermions without hypercharge (see-saw type III~\cite{Foot:1988aq,Ma:2002pf}). Both, type I and type III see-saws, contain only one new physics scale, the Majorana mass of the new fermions, $M$, and lead to the same see-saw formula. Therefore, to explain the observed tiny neutrino masses one needs an extremely large $M$ making very difficult to test these mechanisms\footnote{This conclusion can be avoided if a larger number of fermions is introduced with a particular structure of the mass matrix containing different scales, as in the so-called inverse see-saw mechanism\cite{Wyler:1982dd,Mohapatra:1986bd}, or taking very small Yukawa couplings as in the case of Dirac neutrinos.}. 

Alternatively, to generate neutrino masses at tree level one can enlarge the SM with a scalar triplet with hypercharge $Y = 1$, which develops a VEV (see-saw type II~\cite{Konetschny:1977bn,Cheng:1980qt,Lazarides:1980nt,Magg:1980ut,
Schechter:1980gr,Mohapatra:1980yp}). The model contains two mass scales, the mass of the new particles, $M$, and a trilinear coupling which breaks explicitly LN, $\mu$. Then, the neutrino masses are  $m_{\nu}\sim \mu v^2/M^2$. If $\mu\sim M$ we are in a situation similar to see-saws type I and III, the scale must be very large and the model difficult to test. However, since $\mu$ is protected by symmetry (it is the only coupling in the Lagrangian that breaks LN), it can be naturally small, $\mu \ll M$. Then, $M$ can be at the electroweak scale and the new particles could be produced at the LHC and/or give large effects  
effects in low energy experiments by virtual exchange (for instance, lepton flavour violating processes).  

In the see-saws one explains the smallness of neutrino masses by introducing large mass scales as compared to the electroweak scale. The new heavy particles give huge loop contributions to the SM Higgs mass, which show in all its crudity the hierarchy problem of the SM. This problem can be alleviated in models in which neutrino masses are suppressed by loop factors. For instance, if we enlarge the SM with only scalars, other than triplets with $Y=1$, neutrino masses cannot arise at tree level. However, if LN is broken in the scalar potential, sooner or later Majorana neutrino masses will appear as radiative corrections. Typically this mechanism gives a neutrino mass formula like the see-saw type II, but suppressed by loop factors, $m_{\nu}\sim \frac{1}{(16\pi^2)^n}\mu v^2/M^2$, where $n=1,2,3,\cdots$ is the number of loops at which the mass is generated and $\mu$ is the coupling of the potential that breaks LN. Moreover, the breaking of LN involves the simultaneous presence of several Yukawa couplings which produces further suppressions in the neutrino mass formula. With all these suppressions, the mass of the new particles, $M$, can be rather low even if $\mu$ is not small as compared with $M$. This makes these models testable in present and near future experiments. 

The simplest of these models is the Zee-Babu model~\cite{Zee:1985id,Babu:1988ki}, which contains only two complex scalar singlets, singly  and doubly charged, which we will denote as $h$ and $k$ respectively, and gives neutrino masses at two-loops. The model has a very rich phenomenology that has been widely studied, see for instance \cite{Babu:2002uu,AristizabalSierra:2006gb,Nebot:2007bc,Schmidt:2014zoa,Herrero-Garcia:2014hfa}. The Zee-Babu model is just a representative of a large class of interesting models which give small radiative neutrino masses by extending only the Higgs sector. Archetypes of this class of models are the Zee model \cite{Zee:1980ai} for masses generated at one loop (see \cite{AristizabalSierra:2007nf} for one-loop models with leptoquarks), refs. ~\cite{Zee:1985id,Babu:1988ki,Chen:2006vn,delAguila:2011gr} for two-loop masses (see \cite{Babu:2010vp} for a model with leptoquarks) and \cite{Gustafsson:2012vj,Alcaide:2017xoe} for three-loop masses (see also \cite{Cai:2017jrq} for a recent review and a complete list of references). These models are further motivated by the fact that, contrary to the rest of the SM interactions, departures from the SM can be plausibly hidden in the scalar sector, which is not precisely measured yet. Most of the models we shall be interested contain doubly-charged scalars~\footnote{One prominent exception is the Zee model~\cite{Zee:1980ai}, which contains only singly-charged scalars.}, which have a very rich and peculiar phenomenology.

Several experimental searches for doubly-charged scalars have been carried out at the LHC~\cite{Aad:2011vj,Aad:2012xsa,ATLAS:2012hi,Chatrchyan:2012ya,ATLAS:2016pbt,CMS:2017pet}. They all concluded with negative results. However, unlike some widespread sociological feelings, these results should not be discouraging. On the contrary, a critical assessment of these analyses reveals that they are all not sensitive to doubly-charged scalars with decays other than into same-sign leptons. Departures from this assumption have been already considered in the literature. Thus, pair-produced doubly-charged scalars decaying into $W$ bosons~\cite{Kanemura:2013vxa,Englert:2016ktc} or with both $W$ boson and leptonic decays~\cite{delAguila:2013mia} have been studied. However, no LHC study of doubly-charged scalars with exotic decays, as those arising in models of radiatively-induced neutrino masses (\textit{e.g.} the Zee-Babu model), which are the ones that \textit{must} really have TeV masses, has been worked out so far. In fact, the Zee-Babu model contains the coupling $\mu k^{++}h^- h^-$, which is essential in the generation of neutrino masses and can lead to the decay $k^{\pm\pm}\rightarrow h^\pm h^\pm$. The aim of this paper is to make progress in this direction.

With this spirit, in section~\ref{sec:exotic} we motivate several doubly-charged scalar exotic decay modes that will be subsequently studied in section~\ref{sec:search}. In this section we highlight the most promising LHC observables and signal regions defined out of them to test doubly-charged scalars in a variety of realistic models of neutrino masses. Given the technical difficulties of determining the SM background with good accuracy~\footnote{As we emphasize further in section~\ref{sec:search}, three main challenges affect the generation of the dominant SM backgrounds leading to multi-lepton final states. \textit{(i)} Most of the SM processes contain several particles in the final state. \textit{(ii)} A large fraction of the background is due to charge miss-identification of electrons and positrons coming from huge SM processes such as $Z$+jets. \textit{(iii)} Reaching the TeV region of mass distributions (and other observables with energy dimensions) suggest that NLO-accurate computations must be performed.}, and the fact that our proposed analysis can be eventually used to study prospects for many other scenarios, we also provide background estimation for each of the signal categories considered in this work in appendix~\ref{sec:app1}. In section~\ref{sec:search}, we investigate the reach of this search for the famous Zee-Babu model~\cite{Zee:1985id,Babu:1988ki}, in (very broad) parameter space regions not yet constrained by neutrino data. Finally, we conclude in section~\ref{sec:conclusions}.

\section{Exotic decays of doubly-charged scalars}\label{sec:exotic}
\begin{figure*}[t]
\begin{center}
 \includegraphics[width=0.6\columnwidth]{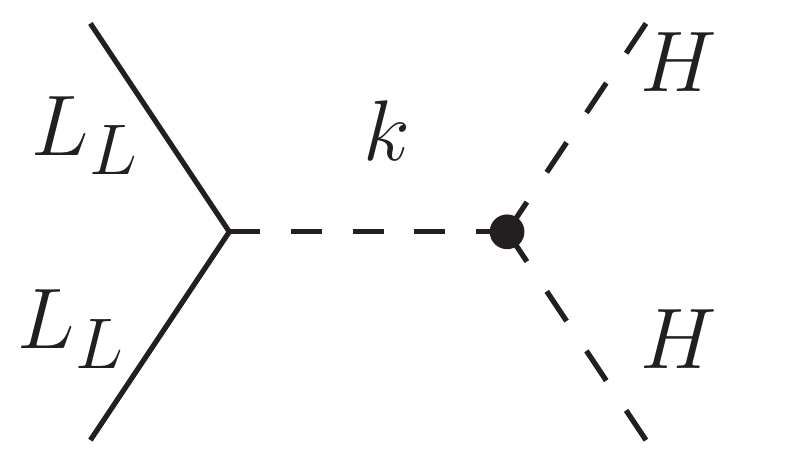}
 \includegraphics[width=0.71\columnwidth]{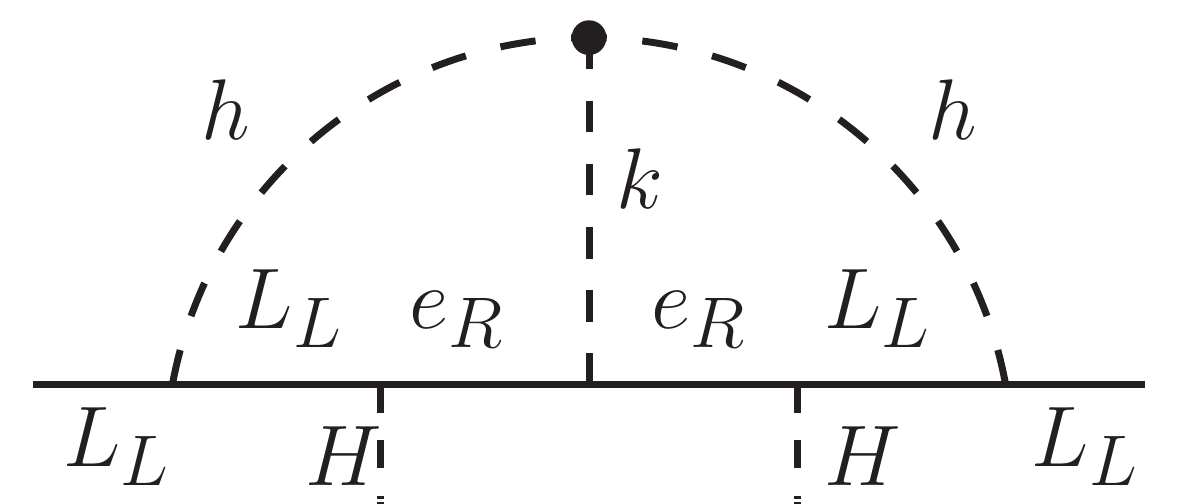}
 \includegraphics[width=0.71\columnwidth]{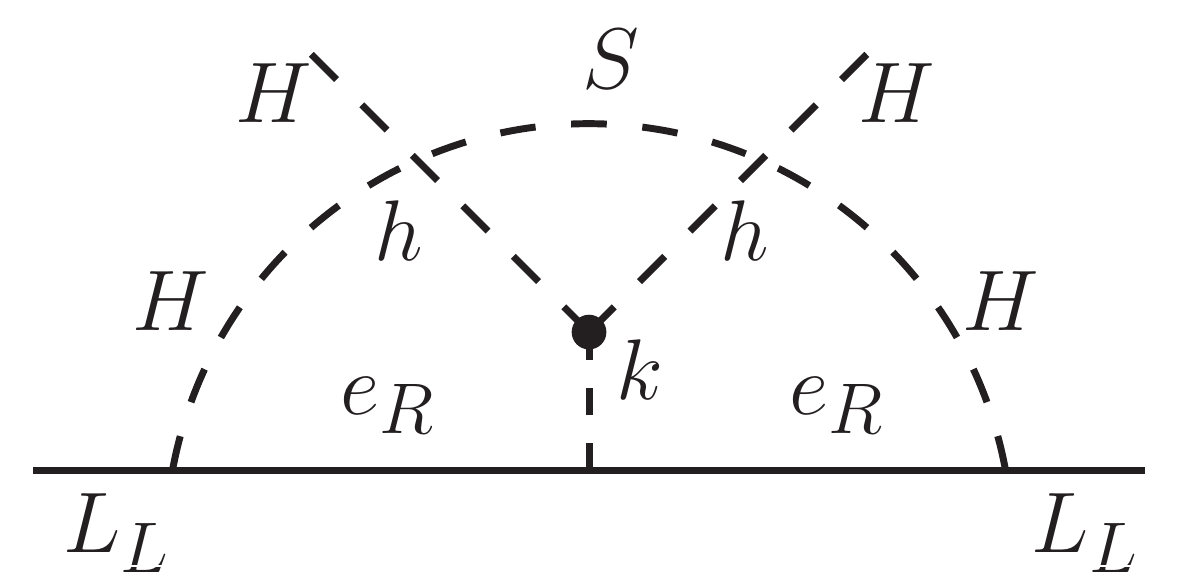}
\end{center}
    \caption{\it Feynman diagrams inducing Majorana neutrino masses in the electroweak phase in the see-saw type II (left), the Zee-Babu model (center) and the model of reference~\cite{Alcaide:2017xoe} (right). $H$ stands for the Higgs doublet. The black dot shows that, provided it is kinematically allowed, $k$ decays into final states other than same-sign leptons (the neutrino masses would vanish otherwise).}\label{fig:diagrams}
\end{figure*}
We will restrict ourselves to SM extensions with only uncolored scalars~\footnote{We remark that all colored scalars with renormalizable couplings to SM fields are flavor-violating, and hence severely constrained; see \textit{e.g.} reference~\cite{deBlas:2014mba}.} with electric charged, at most, $Q = 2$.
We will denote by $k$ the doubly-charged scalar whose decays we are interested in. In addition, we will call $\chi$, $h$ and $S$ any additional doubly-charged, singly-charged or neutral scalar, respectively. Besides the leptonic decay of $k$, $k^{\pm\pm}\rightarrow\ell^\pm\ell^\pm$,
new decay modes are typically present. In fact, $k$ must also couple linearly to non-leptonic fields, because otherwise it would not break LN and therefore would not generate (LN violating Majorana) masses for the neutrinos. (Note that Dirac masses can not be induced by purely scalar extensions of the SM.) In particular, the following $k$ decay modes take place in a variety of models:

$\mathbf{k^{\pm\pm}\rightarrow W^\pm W^\pm}$. It appears even in the simplest model, the see-saw type II, which extends the Higgs sector with an $SU(2)_L$ triplet with $Y=1$; see the left panel of figure~\ref{fig:diagrams} (the $W$ bosons are hidden in the longitudinal components of the Higgs doublet). Provided the
VEV of the neutral component of this triplet is large enough, $k$ decays predominately into gauge bosons. As an example, for a $k$ mass $m_k\sim 500$ GeV, and assuming the neutrino masses to fulfill $\sum m_{\nu_i}^2 = 0.1^2$ eV$^2$, the triplet VEV has to be only above $\sim 0.0001$ GeV~\cite{delAguila:2013aga}. (Note that this value does not spoil the $\rho$ parameter bound on this VEV, $\lesssim$ few GeV.)

Finally, this decay mode appears also naturally in extended composite Higgs models~\cite{Englert:2016ktc}, which are further motivated by the gauge-hierarchy problem.

$\mathbf{k^{\pm\pm}\rightarrow h^\pm h^\pm, h^\pm\rightarrow\ell^\pm \nu}$. Most importantly, it  is the only other possible $k$ decay that occurs in the Zee-Babu; see the center panel of figure~\ref{fig:diagrams}. %

$\mathbf{k^{\pm\pm}\rightarrow h^\pm h^\pm, h^\pm\rightarrow W^\pm S}$. It has been shown to occur in models where both $h$ and $S$ are odd under a $\mathbb{Z}_2$ symmetry, while $k$ is even; see $\textit{e.g.}$ reference~\cite{Gustafsson:2012vj}. A more recent example is given by the model of reference~\cite{Alcaide:2017xoe}; see figure~\ref{fig:diagrams} right panel. Contrary to the first one, in this model $h$ is part of an $SU(2)_L$ triplet with $Y=1$ (instead of a doublet with $Y=1/2$) containing an additional doubly-charged scalar $\chi$. Consequently, we can also find the following decay mode:

$\mathbf{\chi^{\pm\pm}\rightarrow k^{\pm\pm} S, k^{\pm\pm}\rightarrow \ell^\pm\ell^\pm}$. An interesting aspect of this channel is that, for given $\chi$ and $S$ masses, it can sensibly \textit{strengthen} the bounds on $m_k$ coming from current searches for doubly-charged scalars. 

Finally, despite being potentially present, we do not consider cascade decays of $k$ via emission of $W$ bosons. As we will comment below, these are very hard to constrain experimentally.

\section{Search strategy}\label{sec:search}
\begin{figure}[t]
 \begin{center}
  \includegraphics[width=\columnwidth]{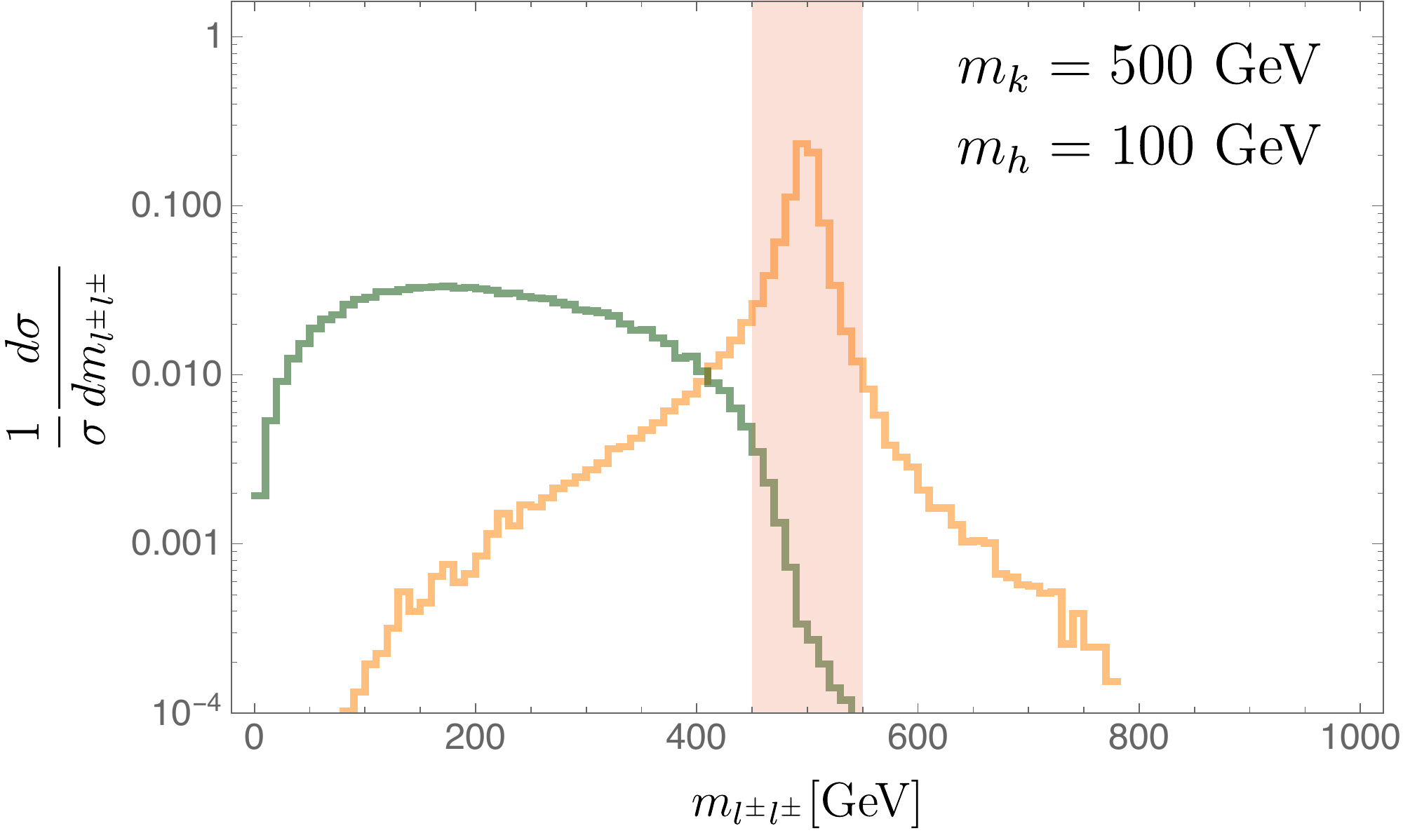}
 \end{center}
\caption{\it Orange: Invariant mass distribution of each pair of two same-sign leptons in events $p p \rightarrow k^{++} k^{--}$ with $k^{\pm\pm}\rightarrow \ell^\pm\ell^\pm$. Green: Same as before but with $k^{\pm\pm}\to h^{\pm}h^\pm, h^\pm\to\ell^\pm\nu$. Red: cut imposed by the experimental collaboration~\cite{CMS:2017pet}. 
$m_k$ and $m_h$ have been set to 500 and 100 GeV, respectively.}\label{fig:comparison1}
\end{figure}

\begin{figure}[t]
 \begin{center}
  \includegraphics[width=\columnwidth]{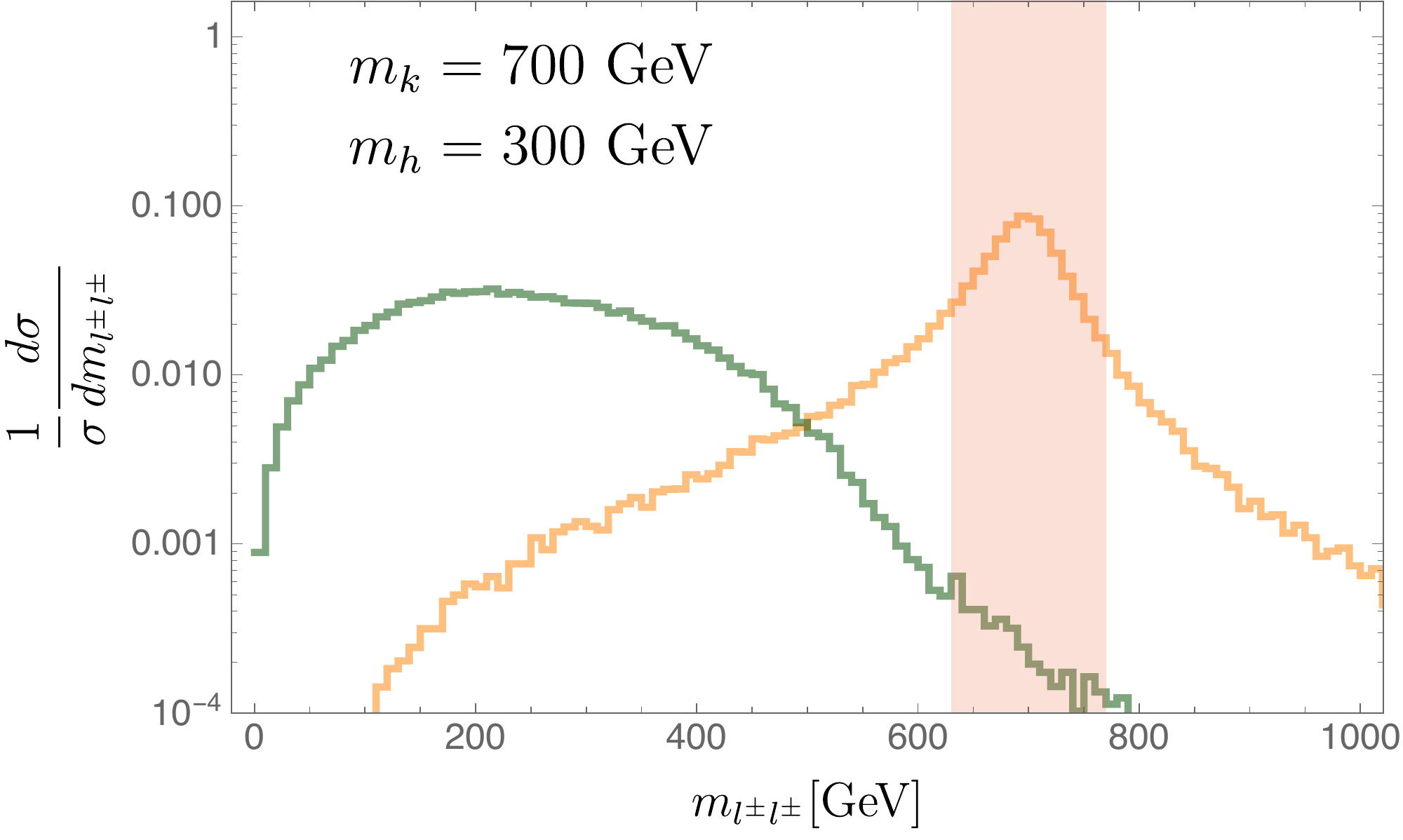}
 \end{center}
\caption{\it Same as figure~\ref{fig:comparison1} but for $m_k = 700$ GeV and $m_h = 300$ GeV.}\label{fig:comparison2}
\end{figure}

Both the ATLAS and the CMS collaborations have developed a large amount of searches for doubly-charged scalars. These include analyses of the total collected luminosity at 7~\cite{Aad:2012xsa,ATLAS:2012hi}, 8~\cite{ATLAS:2014kca,CMS:2016cpz} and 13~\cite{ATLAS:2016pbt,CMS:2017pet,ATLAS:2017iqw} TeV of center of mass energy. However, they are all inspired by the see-saw  type II and therefore look for final states containing pairs of same-sign leptons reconstructing a narrow invariant mass. This requirement is only (and slightly) relaxed in final states with taus. Consequently, doubly-charged scalars with exotic decays can easily be missed. As an example, we compare in figures~\ref{fig:comparison1} and~\ref{fig:comparison2} the invariant mass distribution of the two same-sign lepton pairs resulting from the decay of doubly-charged scalars into $\ell^\pm\ell^\pm$ (orange) and $h^\pm h^\pm$ with $h^\pm\to\ell^\pm\nu$ (green). Clearly, the narrow cut removes most of the signal in this latter case.

Moreover, the interplay between different variables such as the invariant mass of pairs of leptons or the missing energy is never fully exploited. In addition, other analyses, and in particular searches for Supersymmetry in multi-lepton final states, are also non constraining (even for small doubly-charged scalar masses). We have tested this by means of \texttt{CheckMATE v2}~\cite{Dercks:2016npn}, which implements, among others, searches for gluinos in final states with 2 same-sign leptons or 3 leptons, jets and missing energy~\cite{Aad:2016tuk}.
A last analysis that might be sensitive to the signals we are interested in is given by the CMS search for the seesaw type-III in multi-lepton final states~\cite{Sirunyan:2017qkz}. Again, this search is of narrow scope. Among the characteristics that make it not suitable to explore generic doubly-charged scalars we find that it focuses on final states with same-flavour opposite-sign leptons. For models such as the Zee-Babu, where the doubly-charged scalar can produce equally muons or electrons, this requirement kill half of the signal. Furthermore, final states with only two light leptons, which are abundant in the models of interest, are disregarded. In any case, that CMS paper does not provide detailed information (\textit{e.g} number of background events in each signal category), and so a proper estimation of its (presumably limited) reach to these models can not be precisely stated. To the best of our knowledge, no other multi-lepton analysis with the latest data has been made public yet.

As things stand, new analyses are necessary to fully explore models of radiatively-induced neutrino masses. With the aim of being able to test different scenarios (though still focusing on neutrino models~\footnote{Despite being interesting, inclusive searches looking ``everywhere'' \textit{without} any theoretical biased, such as the model-independent analysis of reference~\cite{ATLAS:2014sxa}, are not very efficient in the search for new physics.}), we propose a broad scope search containing several signal regions and categories.

They all contain several same-sign leptons. One of the main challenges of this kind of analysis is the correct estimation of the background, which originates mainly from the charge miss-identification of electrons. This requires time-consuming simulations of SM processes with many particles in the final state. Consequently, given the numerous signal regions that we work out in the next section, our background estimation will be valuable for many LHC studies of particles producing same-sign lepton events.

\subsection{New signal regions}
Prior to the selection of events, the relevant physical objects are constructed in the following way. Electrons (muons) are defined to have $p_T^\ell > 20~(10)$ GeV and $|\eta_\ell| < 2.5~(2.6)$. Jets are clustered using the anti-$k_t$ algorithm with $R = 0.4$. They are defined by $p_T^j > 20$ GeV and $|\eta_j| < 2.4$. Despite their small phenomenological relevance, we have also fixed the $b$-tagging efficiency to 0.7 and the $\tau$-tagging efficiency to 0.5. Of major importance is the probability of an electron (positron) to be identified as a positron (electron). Following reference~\cite{ATLAS:2016pbt} 
we estimate it by $P(|\eta_\ell|, p_T^\ell) = f(|\eta_\ell|)\sigma(p_T^\ell)$ with
\begin{equation}\label{eq:emisstag1}
 f(x) = \left\lbrace\begin{array}{lll}
                             0.03 & \text{if} & 0<x<0.4,\\
                             0.04 & \text{if} & 0.4<x<0.8,\\
                             0.08 & \text{if} & 0.8<x<1.1,\\
                             0.15 & \text{if} & 0.8<x<1.1,\\
                             0.3 & \text{if} & 1.1<x<1.4,\\
                             0.6 & \text{if} & 1.4<x<1.7,\\
                             0.7 & \text{if} & 1.7<x<1.9,\\
                             1 & \text{if} & 2.1<x<2.3,\\
                             2 & \text{if} & x> 2.3.
                            \end{array}\right.
\end{equation}
and
\begin{equation}\label{eq:emisstag2}
 \sigma(x) = \left\lbrace\begin{array}{lll}
										0.02 & \text{if} & x<70,\\
										0.035 & \text{if} & 70<x<100,\\
										0.05 & \text{if} & x>100.\\
										\end{array}\right.
\end{equation}
Throughout the text, we will refer to electrons and muons simply as \textit{leptons}, while taus will be excluded of this definition.
Only events with at least two-same sign leptons are selected. Out of this sample, we define three orthogonal signal regions (SRs), containing two, three and four leptons, respectively.

\textbf{SR 1}: inspired by the recent ATLAS analysis of reference~\cite{ATLAS:2016pbt}, 
it contains events with two leptons. If more than two same-sign leptons are present, only that pair with the highest invariant mass is considered for computing the observables defined below.

\textbf{SR 2}: inspired by the recent CMS analysis of reference~\cite{CMS:2017pet},  
it contains events with exactly three leptons, with exactly two of opposite sign.

\textbf{SR 3}: inspired by the same CMS analysis, it contains events with exactly two positive and two negative leptons.

We further consider the following observables. 1) $S_T$, defined as the scalar sum of the $p_T$ of all leptons in the signal region. 2) The invariant mass of each same-sign lepton pair, $m_{\ell^\pm\ell^\pm}$. 3) The transverse mass of each same-sign lepton pair, as well as the one of the third lepton in SR 3. We will denote both  collectively by $m_T$. Following reference~\cite{Englert:2016ktc}, we define the former by
\begin{align}
 m_T^2 = &\left[\sqrt{(p_T^{\ell^\pm\ell^\pm})^2 + m_{\ell^\pm\ell^\pm}^2} + E_T^{\text{miss}}\right]^2 \\
 &- \left[p^{\ell^\pm\ell^\pm}_x  + E_x^{\text{miss}}\right]^2 - \left[p^{\ell^\pm\ell^\pm}_y + E_y^{\text{miss}}\right]^2~.
\end{align}
where $E^{\text{miss}}$ stands for the missing energy. 4) The stransverse mass, $m_{T2}$, defined as
\begin{align}
 m_{T2} = \text{min}_{\mathbf{q}_T}\bigg\lbrace\text{max}\bigg[p_T^{L_1} E_T^{\text{miss}}-\mathbf{p}_T^{L_1}\cdot \mathbf{q}_T,\\
 p_T^{L_2} E_T^{\text{miss}}-\mathbf{p}_T^{L_2}\cdot (\mathbf{E}_T^{\text{miss}}-\mathbf{q}_T)\bigg]\bigg\rbrace~.
\end{align}
In SR1, $L_1$ and $L_2$ are given by the harder and the softer lepton, respectively. In SR2, $L_1$ stands for the vectorial sum of the two same-sign leptons, while $L_2$ is given by the third one. In SR3, $L_1$ represents the vectorial sum of the two positive charged leptons, and $L_2$ the vectorial sum of the two negative ones.

For each SR, and for each observable $O = $ $m_{\ell^\pm\ell^\pm}$, $m_T$, $m_{T2}$, we consider $81$ different categories defined by $S_T > X$ and $O > Y$ with $X, Y = 100, 200, ...,  900$ GeV.
\begin{table}[t]
 \begin{center}
\begin{adjustbox}{width=0.95\columnwidth}
\footnotesize
\begin{tabular}{|c|c|c|}\hline
              &  &  \\[-0.2cm]
              &  ~~~~~~$\sigma$ [pb]~~~~~~ & ~\# MC events~ \\[0.1cm]\hline
   ~~Drell-Yan~~           &        $220\pm 20$      & $10^8$\\
   $t\overline{t}$          &    $660\pm 70$          & $10^8$\\
   $WW$           &      $102\pm 4$        & $10^7$\\
   $WZ$           &       $45\pm 2$       & $10^6$\\
   $ZZ$           &      $13.6\pm 0.5$        & $10^6$\\
   $WWW$           &       $0.21\pm 0.01$       & $10^6$\\
   $WWZ$           &       $0.17\pm 0.01$       & $10^6$\\
   $WZZ$           &       $0.057\pm 0.004$       & $10^6$\\
   $ZZZ$           &       $0.014\pm 0.001$       & $10^6$\\
   $t\overline{t} W$           &        $0.59\pm 0.06$      & $10^6$\\
   $t\overline{t} Z$           &     $0.76\pm0.09$         & $10^6$\\
 \hline
 \end{tabular}
 \end{adjustbox}
 \end{center}
 \caption{\it Backgrounds, cross sections and numbers of generated Monte Carlo events.}\label{tab:backgrounds}
\end{table}
In accord with reference~\cite{CMS:2017pet}, %
we consider the following dominant backgrounds: Drell-Yan with $m_{\ell^+ \ell^-} > 100$ GeV, $t\overline{t}$, $WZ$, $WW$, $ZZ$, $WWW$, $WWZ$, $WZZ$, $ZZZ$, $t\overline{t}W$, $t\overline{t}Z$. Background events are generated at NLO in $\alpha_s$ with \texttt{MadGraph v5}~\cite{Alwall:2014hca}. Initial and final state radiation and showering is performed by \texttt{Pythia v6}~\cite{Sjostrand:2006za}. The cross sections of all relevant backgrounds are shown in table~\ref{tab:backgrounds}. The uncertainty due to the choice of scale is also shown. Finally, we also provide the number of generated Monte Carlo events.

\begin{figure}[t]
\begin{center}
 \includegraphics[width=\columnwidth]{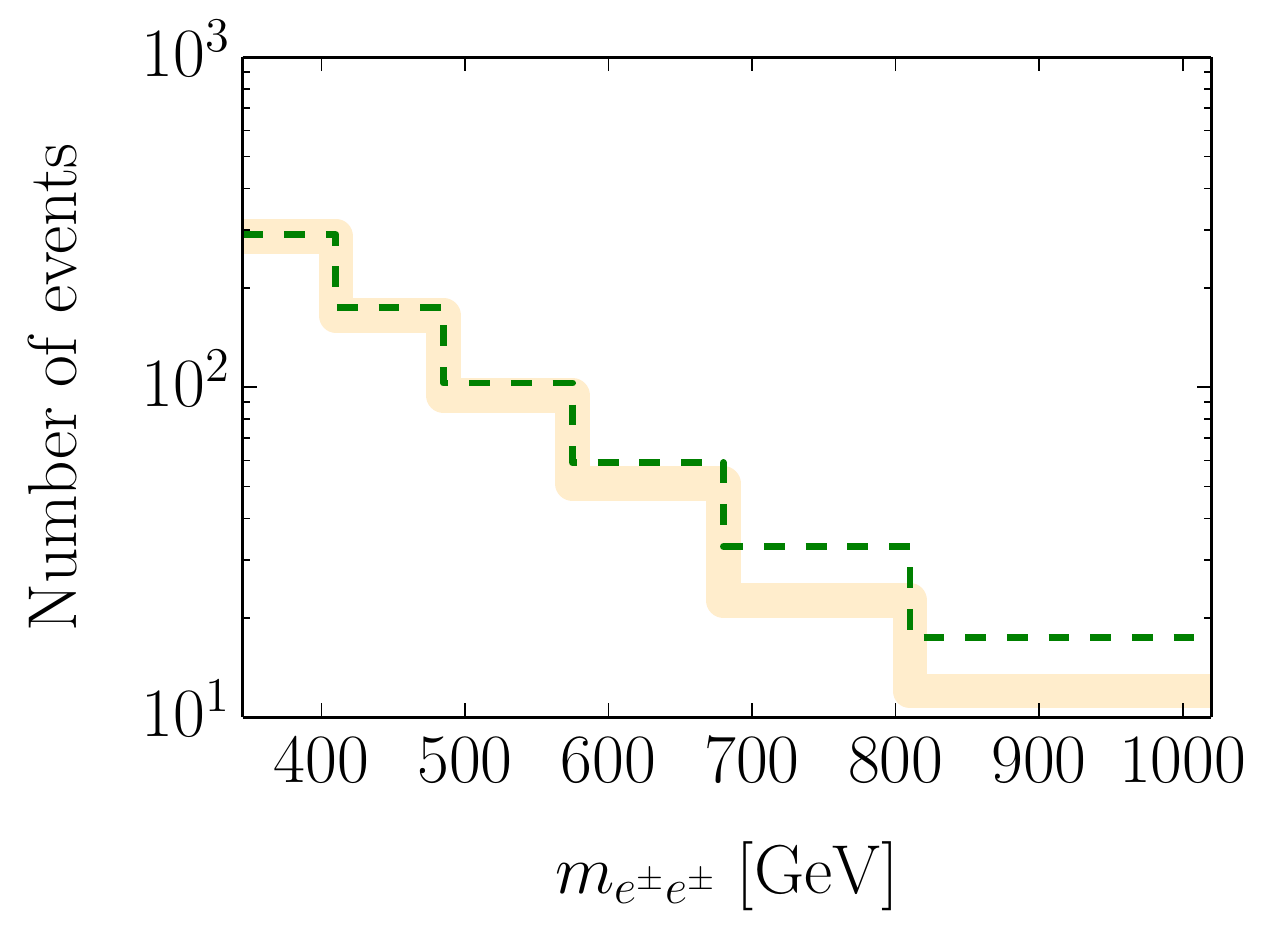}
\end{center}
\caption{\it Number of events with $m_{e^\pm e^\pm} >$ $345,$ $410,$ $485,$ $575,$ $680,$ $810,$ $1020$ GeV after the preselection cuts of reference~\cite{ATLAS:2016pbt}. 
The orange solid line stands for our result, while the green dashed one corresponds to the results provided by the ATLAS collaboration. The small discrepancy at large invariant masses is not relevant in practice, because the SM background is almost negligible in that region.}\label{fig:validation1}
\end{figure}

In order to validate the goodness of our Monte Carlo generation as well as the appropriateness of equations~\ref{eq:emisstag1} and \ref{eq:emisstag2}, we recast the analysis of reference~\cite{ATLAS:2016pbt} %
and compare the distribution of $m_{e^\pm e^\pm} > 345, 410, 485, 575, 680, 810$ and $1020$ GeV provided by the experimental collaboration with that obtained by us. For this goal, we use homemade routines based on \texttt{MadAnalysis v5}~\cite{Conte:2012fm} and \texttt{ROOT}~\cite{Brun:1997pa}. The result is depicted in figure~\ref{fig:validation1}.

\begin{figure}[t]
\begin{center}
 \includegraphics[width=1.042\columnwidth]{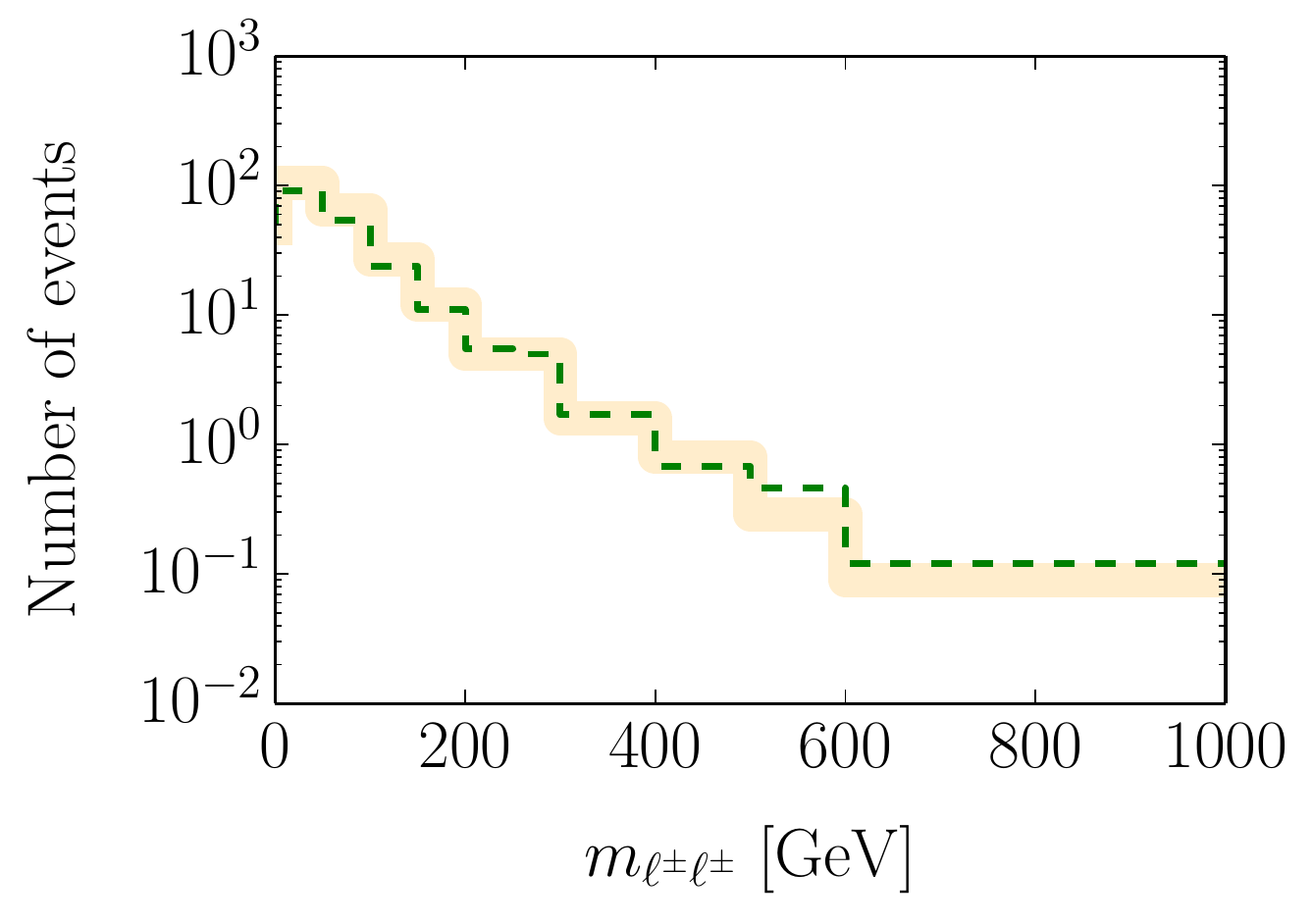}
\end{center}
\caption{\it Event distribution for $m_{\ell^\pm\ell^\pm}$ after the preselection cuts of reference~\cite{CMS:2017pet}. 
The orange solid line stands for our result, while the green dashed one corresponds to the results provided by the CMS collaboration.}\label{fig:validation2}
\end{figure}

We also compare the distribution of $m_{\ell^\pm\ell^\pm}$ given in reference~\cite{CMS:2017pet} %
with ours; it can be seen in figure~\ref{fig:validation2}. Clearly, our results are in perfect agreement with those provided by the experimental collaborations. Moreover, we have checked that the contribution of each background in table~\ref{tab:backgrounds} to the total SM expectation agrees with that reported by both ATLAS and CMS. Consequently, we have computed the number of expected background events in each of the categories mentioned above. These are listed in tables~\ref{tab:11}, \ref{tab:12}, \ref{tab:13}, \ref{tab:21}, \ref{tab:22}, \ref{tab:23}, \ref{tab:31}, \ref{tab:32} and \ref{tab:33} in the appendix~\ref{sec:app1}.

\subsection{Applications}\label{sec:applications}
Based on this information, we can estimate the reach of current data to several signals mediated by doubly-charged scalars. Let us start considering the standard case $k^{\pm\pm}\to \ell^{\pm}\ell^{\pm}$. This is the only one that has been considered in experimental analyses so far; it will allow us to further validate our approach. We focus on pair-production of doubly-charged scalars. As a matter of fact, this channel is always present, while associated production is absent in many models; see section~\ref{sec:exotic}.

We implement the relevant interactions in \texttt{Feynrules v2}~\cite{Alloul:2013bka}. We generate signal events using \texttt{MadGraph v5} at LO in $\alpha_s$, using again \texttt{Pythia v6} as parton shower. We then estimate the number
of signal events in each of the categories defined above for $\mathcal{L} = 35$ fb$^{-1}$. (We restrict to this value because no experimental analysis with more data is still publicly available.) We subsequently look for those three categories that give the largest sensitivity defined as $S/\sqrt{B}$ in SR1, SR2 and SR3, respectively. (Obviously, in the present four-lepton case, the sensitivity is by far driven by the categories with $m_{\ell^\pm\ell^\pm} \gtrsim m_{k^{\pm\pm}}-100$ GeV in SR3.)

These three categories are orthogonal, meaning that no single event can belong to more than one of them. We can therefore simultaneously consider these three independent categories to analyze to what extent the signal is compatible with the observed data given the SM predictions of appendix~\ref{sec:app1}. To this aim, we adopt the CL$_\text{s}$ method~\cite{Read:2002hq}. 
The corresponding statistic is computed using \texttt{MCLimits}~\cite{Junk:1999kv}, which takes also into account the systematic uncertainty due to the finite number of generated Monte Carlo events. On top of it, we include a systematic uncertainty in the background normalization of $10~\%$. For each value of $m_k = 300, 500, 700, 900$ GeV, we estimate the lowest cross section that can be excluded at the 95 \% C.L. using this procedure. Exclusions for intermediate masses are obtained by linear interpolation.
\begin{figure}[t]
\begin{center}
 \includegraphics[width=1.042\columnwidth]{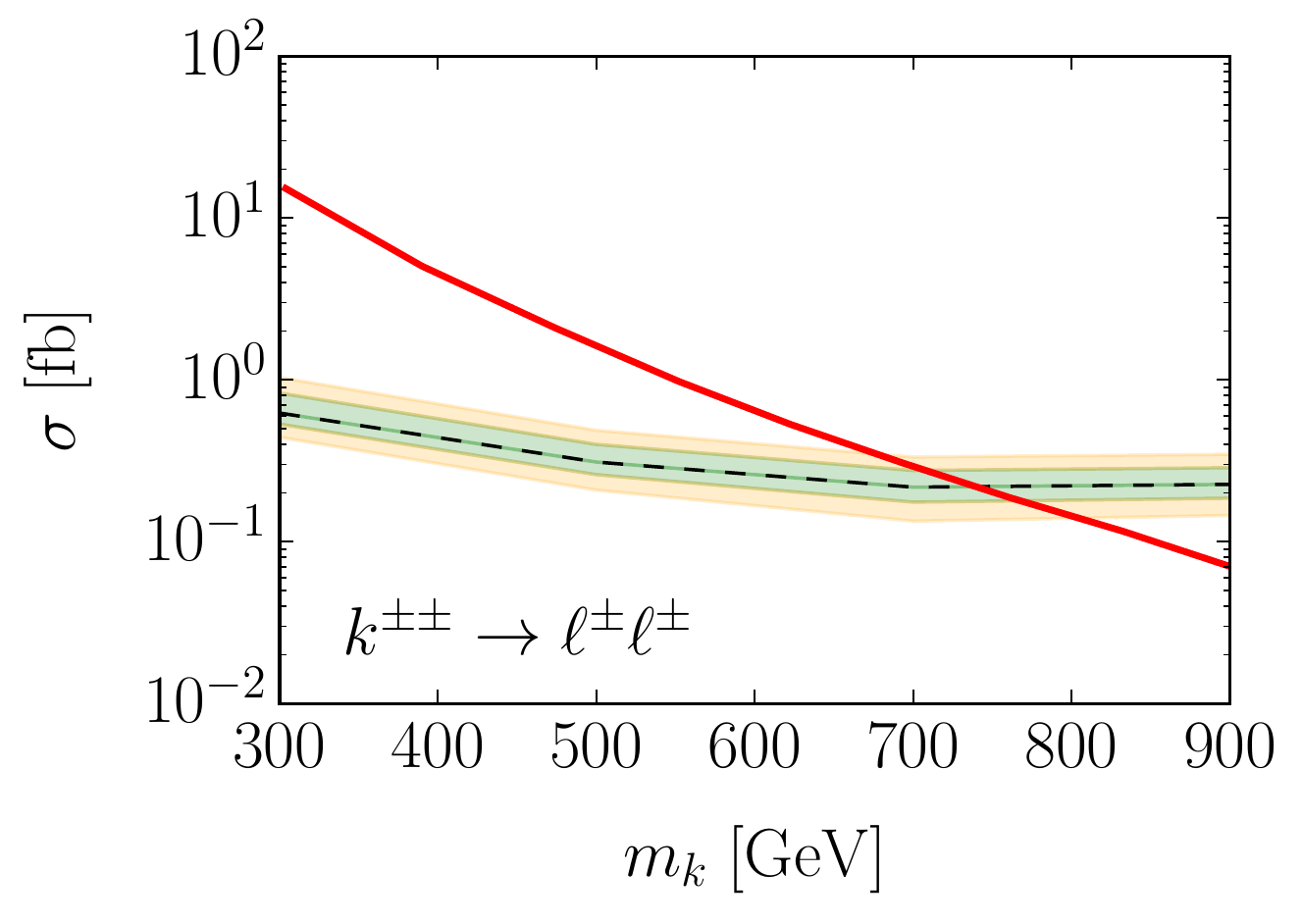}
\end{center}
 \caption{\it Bounds on the cross section of pair-produced $k$ decaying into same-sign leptons (dashed black line). The green and orange regions show the $1~\sigma$ and $2~\sigma$ uncertainties. We have fixed $\mathcal{L} = 35$ fb$^{-1}$. The theoretical cross section in the see-saw type II is also shown for reference (solid red line).}\label{fig:ll}
\end{figure}

The results are depicted by the black dashed line in the left panel of figure~\ref{fig:ll}. The shaded green and orange regions represent the $1\,\sigma$ and the $2\,\sigma$ error bands. For concreteness, we superimpose the theoretical cross section predicted in the see-saw type II. The results are in very good agreement with those presented in the experimental works of references~\cite{ATLAS:2016pbt} %
and \cite{CMS:2017pet}. %
As things stand, masses as large as $m_{k}\sim 700$ GeV are excluded in this channel by using current data. Note however that these are significantly weakened in models in which $k$ is an $SU(2)_L$ singlet, whose production cross section is roughly a factor of $2$ smaller.

One can easily derive approximate prospects for a larger luminosity, by scaling the cross section bound by $\sqrt{35~ \text{fb}^{-1}/\mathcal{L}}$. Thus, the cross section limit at large masses goes down one order of magnitude for $\mathcal{L} = 3$ ab$^{-1}$. The upper limit on the triplet masses turns out to be in this case $\sim 1.1$ TeV. We note, however, that some corrections to this result might be needed, given the limited statistic of some Monte Carlo samples (see table~\ref{tab:backgrounds}) for large luminosities.

\begin{figure}[]
\begin{center}
  \includegraphics[width=0.95\columnwidth]{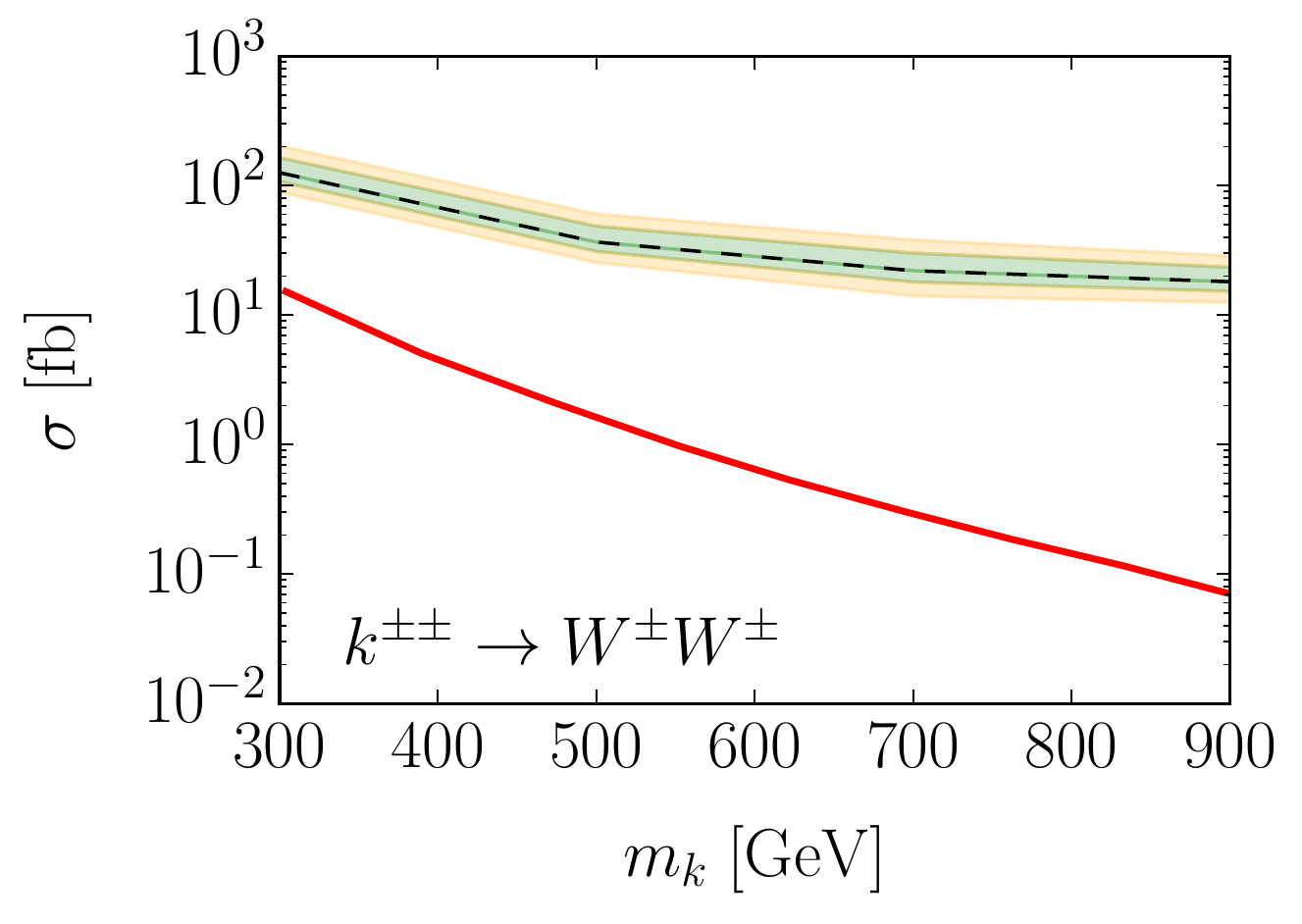}
\end{center}
 \caption{\it Same as figure~\ref{fig:ll} but for $k^{\pm\pm}\to W^\pm W^\pm$.}\label{fig:ww}
\end{figure}

We repeat this exercise for the different signals commented in section~\ref{sec:exotic}, to which current analyses, relying on a narrow cut on the invariant mass of any pair of same-sign leptons, are not sensitive at all. The first such a signal appears when $k^{\pm\pm}\to W^\pm W^\pm$, what can happen also in the see-saw type II. We still restrict to the pair-production mode. (Note that the associated production channel is not necessarily present even in these models with this decay; see for example reference~\cite{Hierro:2016nwm}.) The most sensitive categories are those with $m_T, S_T \gtrsim 100-300$ GeV in SR1, $m_T \gtrsim 100-300$ GeV and $S_T > 400-600$ GeV in SR2 and the ones with $m_{\ell^\pm\ell^\pm} \gtrsim 100-300$ GeV and $S_T \gtrsim 600$ GeV in SR3. In this case, we have checked that the combination of the three categories with two, three and four leptons improve the sensitivity of LHC data by almost an order of magnitude with respect to that obtained using only the most sensitive category. Even so, the presence of $W$ bosons in the final state makes this channel almost unaccessible with current data; see figure~\ref{fig:ww}. In the long term, instead, masses up to $m_{k}\sim 400$ GeV might be probed. Analyses specifically dedicated to this channel in composite Higgs models could improve over this result; see reference~\cite{Englert:2016ktc}.

\begin{figure}[]
\begin{center}
 \includegraphics[width=0.95\columnwidth]{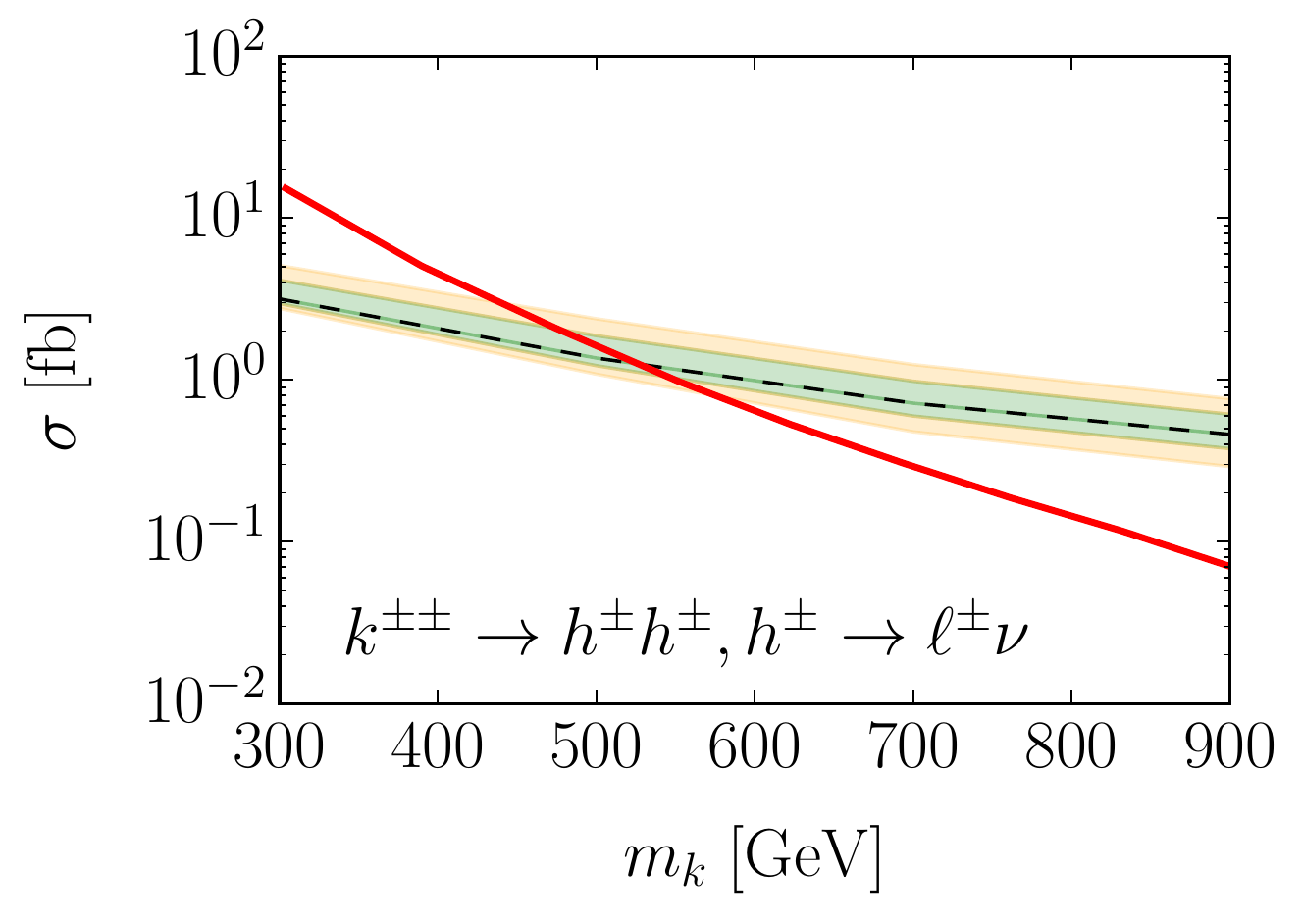}
 \end{center}
 \caption{\it Same as figure~\ref{fig:ll} but for $k^{\pm\pm}\to h^\pm h^\pm, h^\pm\to\ell^\pm \nu$. We fixed $m_{h} = m_{k}/2.5$ as well as $m_{S} = m_{h}/2.5$. Three body decays are considered when necessary.}\label{fig:rr1}
\end{figure}

\begin{figure}[]
\begin{center}
  \includegraphics[width=0.95\columnwidth]{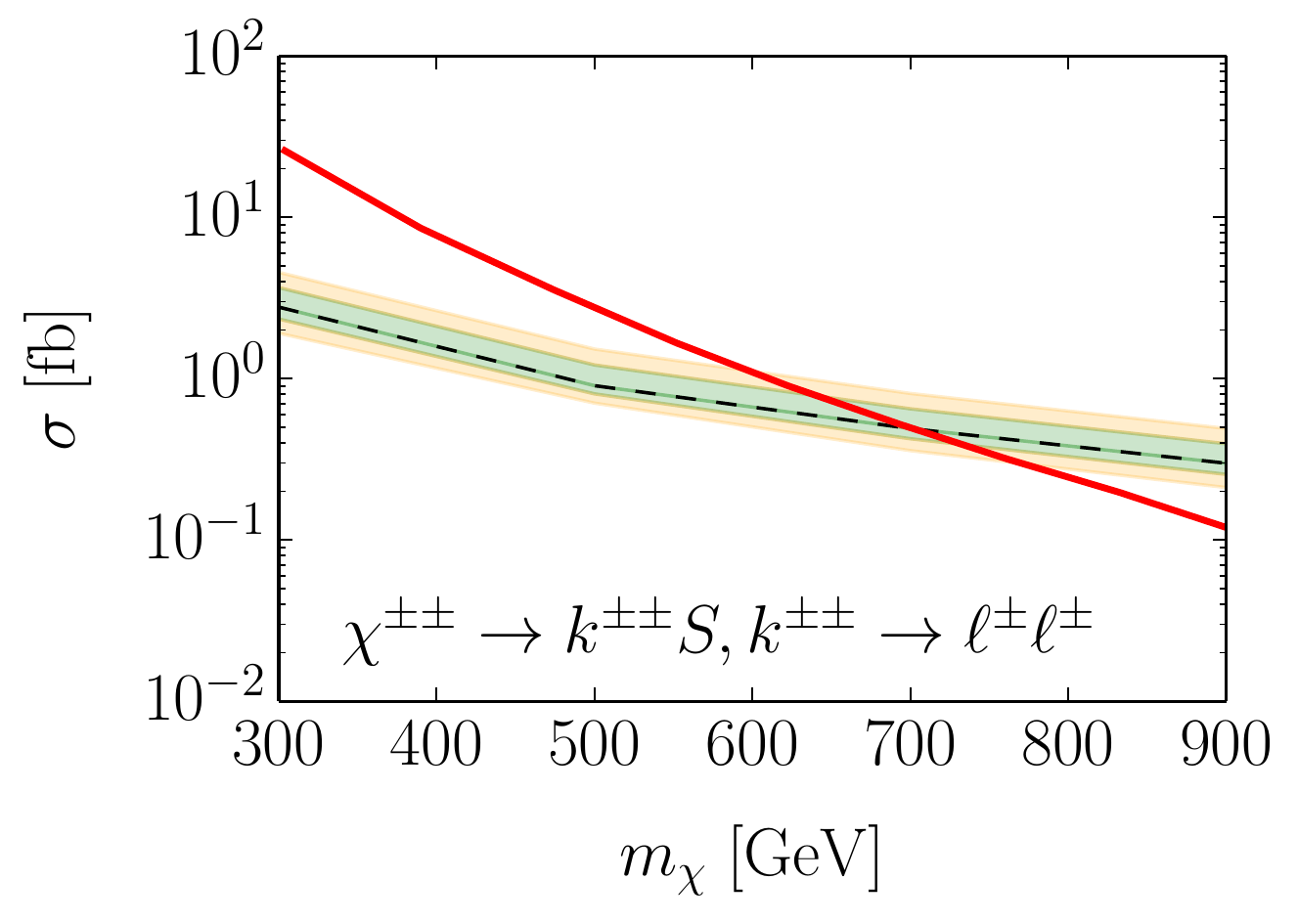}
 \end{center}
 \caption{\it Same as figure~\ref{fig:ll} but for $k^{\pm\pm}\to \chi^{\pm\pm} S, \chi^{\pm\pm}\to\ell^\pm \ell^\pm$. We fix $m_{\chi} = m_{h} = m_{S} = m_{k}/2.5$. Three body decays are considered when necessary.}\label{fig:rr2}
\end{figure}

\begin{figure}[]
\begin{center}
  \includegraphics[width=0.95\columnwidth]{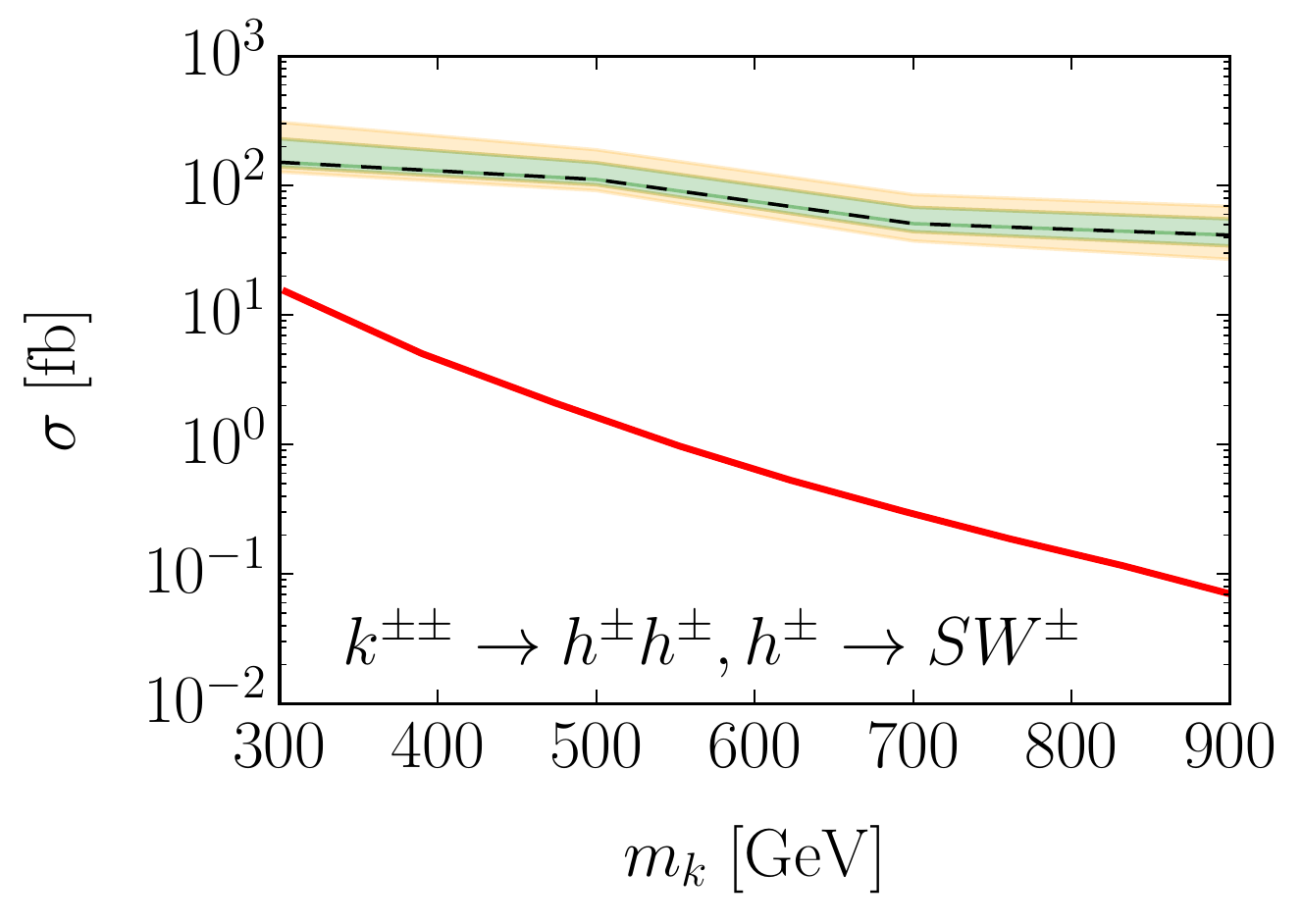}
 \end{center}
 \caption{\it Same as figure~\ref{fig:ll} but for $k^{\pm\pm}\to h^\pm h^\pm, h^\pm\to SW^\pm$. We fix $m_{\chi} = m_{h} = m_{S} = m_{k}/2.5$. Three body decays are considered when necessary.}\label{fig:rr3}
\end{figure}

Using this same broad-scope strategy, we analyze the LHC reach for pair-produced doubly-charged scalars decaying into exotic channels (with $100 \%$ branching ratio). The results are shown in figures~\ref{fig:rr1}, ~\ref{fig:rr2} and \ref{fig:rr3}.
We see that cascade decays with $W$ bosons in the final state are still hard to tag. It can be then easily understood that decay chains with emissions of soft $W$ bosons are also very unconstrained. Therefore, a large parameter space of models giving these decays is perfectly allowed by collider data. On the contrary, the pair-production of $k^{\pm\pm}k^{\mp\mp}$ with the subsequent decay of $k^{\pm\pm}\rightarrow h^{\pm}h^{\pm}$ is very constrained. The most sensitive signal categories in this respect are those with $m_{T2}\sim S_T \gtrsim m_k/2$ in SR1 and those with $m_T\sim S_T \gtrsim m_k/2$ in SR2 and SR3.

\begin{figure}[t]
\begin{center}
 \includegraphics[width=0.95\columnwidth]{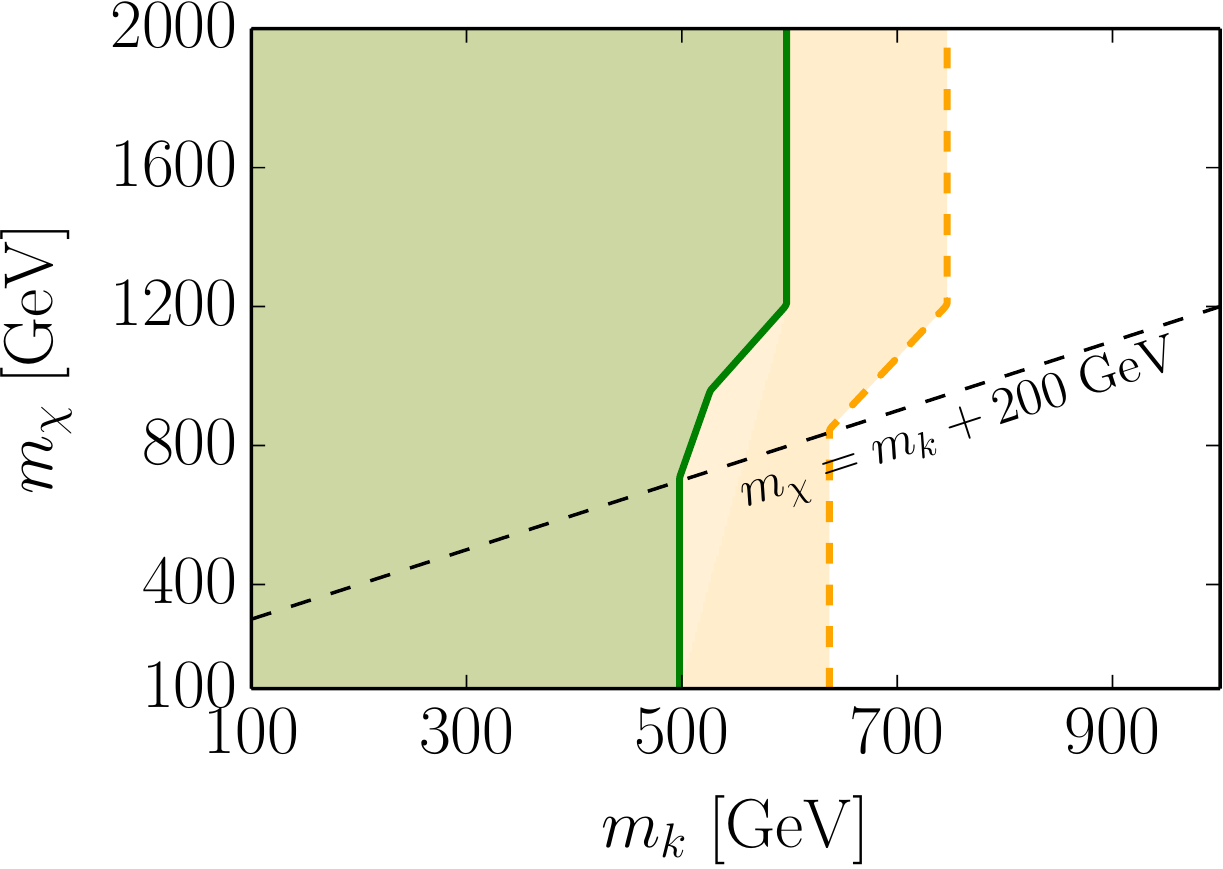}
 \end{center}
 \caption{\it Excluded region for a model containing an $SU(2)_L$ singlet $k^{\pm\pm}$ decaying into $\ell^\pm\ell^\pm$ as well as a doubly-charged component of a $Y=1$ $SU(2)_L$ triplet, $\chi^{\pm\pm}$, decaying mostly into $\kappa^{\pm\pm}$ and a neutral scalar $S$ when kinematically accessible (dashed orange line). The bounds on $m_{k}$ are $\sim 100$ GeV above those obtained assuming the presence of this particle alone. We also show the bounds when $\kappa^{\pm\pm}$ decays mostly into $\ell^\pm\tau^\pm$ (green solid line), as suggested by the recent proposed model of reference~\cite{Alcaide:2017xoe}. For consistency with this model, we have in both cases assummed the mass of $S$ to be 200 GeV.}\label{fig:aamodel}
\end{figure}

Likewise, $\chi^{\pm\pm}\rightarrow k^{\pm\pm} S$ with $S$ stable (\textit{i.e.} missing energy) and $k\rightarrow \ell^\pm\ell^\pm$ is almost as constrained as the standard pair-production of doubly-charged scalars with leptonic decays. Thus, in models in which this production mode is present, constraints on $k$ can be significantly altered with respect to those obtained using standard searches for $k$ alone (see figure~\ref{fig:aamodel}) because the latter is more copiously produced.

\section{Implications for concrete models}
In concrete models, the doubly-charged particles can decay into several different channels, and so the bounds on different parameter space points can not be read from the plots above. Instead, the full process of comparing signal, background and data outlined in section~\ref{sec:applications} must be done. Note, however, that the background, which is the most complicated and time-consuming task in this respect, does not need to be computed again. It can be just taken from the tables in appendix~\ref{sec:app1}. We illustrate this procedure in the Zee-Babu model. It has been previously considered in several references~\cite{Zee:1985id,Babu:1988ki,Babu:2002uu,AristizabalSierra:2006gb,Nebot:2007bc,Schmidt:2014zoa,Herrero-Garcia:2014hfa}. All of them have assumed that the exotic decays of the doubly-charged scalars were impossible to tag at the LHC.

\subsection{The Zee-Babu model}
The Zee-Babu model extends the SM scalar sector with two $SU(2)_L$ singlets, $h$ and $k$, with hypercharges $Y = 1$ and $Y = 2$, respectively. The relevant Lagrangian for our discussion reads
\begin{align}\nonumber
 L = L_{\text{SM}} + f^{ab} \overline{\tilde{L}_{aL}} L_{Lb} h^+ &+ g^{ab}\overline{e^{c}_a} e_b k^{++}\\
 &- \mu k^{++}h^-h^⁻ + \text{h.c.} + \cdots
\end{align}
where $L_{\text{SM}}$ stands for the SM Lagrangian and $L_{aL}$, $\ell_a$ with $a = 1,2,3$ are the first, second and third generation SM lepton doublets and singlets, respectively, and $\tilde{L}_L = i\sigma_2 L_L^c$ with $\sigma_2$ the second Pauli matrix. Overall, the model depends only on the antisymmetric (resp. symmetric) dimensionless couplings $f_{ab}$ (resp. $g_{ab}$), the physical masses of the new scalars, namely $m_k$ and $m_h$ and the dimensionless parameter $\kappa$ defined by $\mu = \kappa~\text{min}\lbrace m_h, m_k\rbrace$. The relevant decay widths read
\begin{equation}
 \Gamma_{k^{\pm\pm}\rightarrow\ell_a^\pm \ell_b^\pm} = \frac{|g_{ab}|^2}{4\pi(1+\delta_{ab})}m_{k}~,
 \end{equation}
 and
 \begin{equation}
 \Gamma_{k^{\pm\pm}\to h^\pm h^\pm} = \frac{1}{8\pi}\left(\frac{\mu}{m_{h}}\right)^2 m_{k}\sqrt{1-\frac{4m_{h}^2}{m_{k}^2}}~.
\end{equation}

Naturalness arguments, together with the requirement that no charge-breaking global minimum is developed by the potential, imply that $\kappa \lesssim 4\pi$. Neutrino oscillation data and low-energy constraints restrict the allowed parameter space. We consider two large regions permitted by current experiments, depending on whether the neutrino mass hierarchy is normal (NH) or inverted (IH)~\cite{Schmidt:2014zoa,Herrero-Garcia:2014hfa}.

\subsubsection{Normal hierarchy}
\begin{figure}[t]
\begin{center}
  \includegraphics[width=0.95\columnwidth]{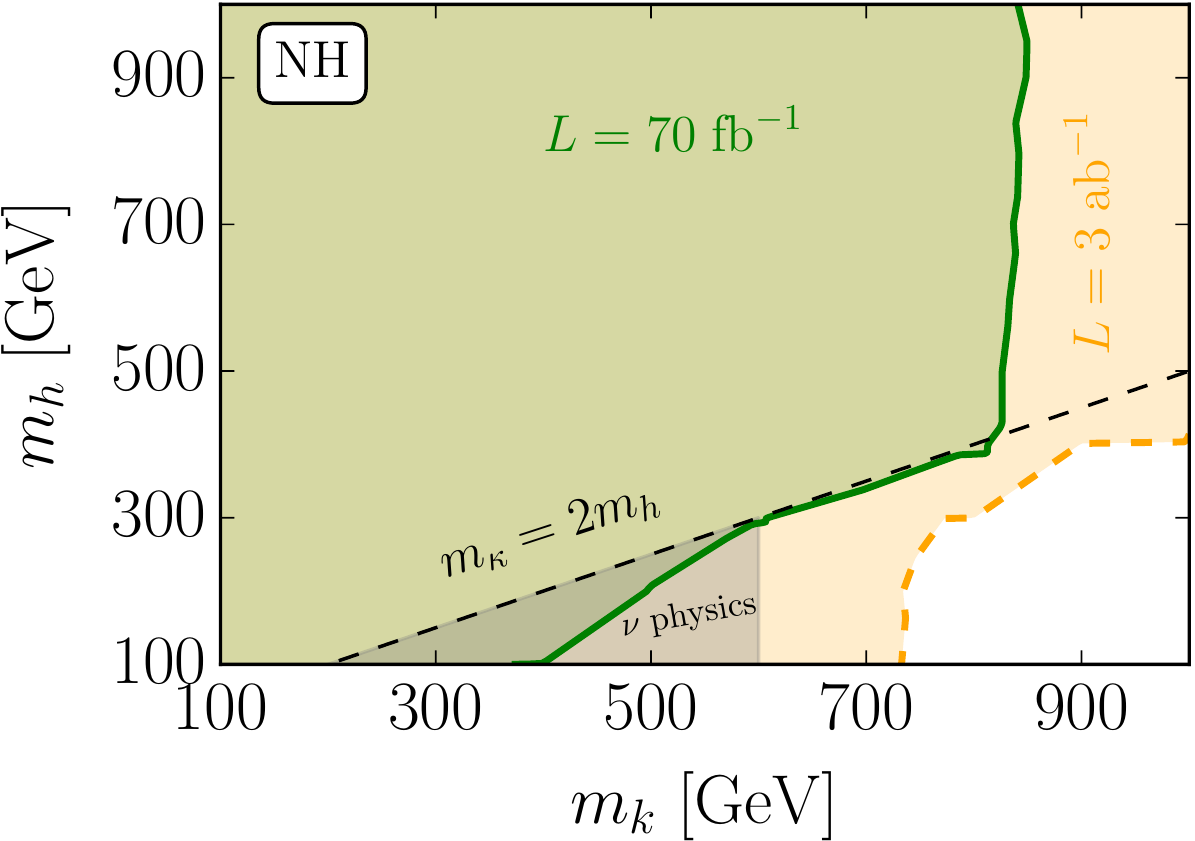}
 \end{center}
 \caption{\it Excluded regions in the plane $m_k-m_h$ in the NH in the Zee-Babu model. See the text for details.}\label{fig:nh}
\end{figure}

According to neutrino data, $g_{11} \sim g_{22}\sim 0.1 \gg$ $g_{12},$ $g_{13},$ $g_{23},$ $g_{33}$. The region $m_k < 2m_h$ is allowed for $m_k > 400$ (resp. $600$) GeV if $\kappa \sim 4\pi$ (resp. $5$). The measured values of the neutrino mixing angles fix $f_{12}\sim f_{13}\sim f_{23}/2$. An overall scale of $f\sim 0.01$ is in agreement with $\mu\to e\gamma$ bounds. Accordingly, we consider the following values:
\begin{align}\nonumber
 g_{11} &= g_{22} = 0.1, g_{12} = g_{13} = g_{33} = 0.001, \\
 f_{12} &= f_{13} = 0.01, f_{23} = 0.02, \kappa = 5. 
\end{align}
As a result, $k$ decays mainly into leptons (for $m_k < 2 m_h$), while $h$ decays into a lepton and a neutrino around 60 \% of times and, to a lesser extent ($\sim$ 40 \%), into a tau and a neutrino. Thus, for $m_k > 2 m_h$, the pair-production of doubly charged scalars give rise to two, three, and four-lepton events in 35 \%, 30 \% and 15 \% of the cases, respectively. Being 0 and 1-lepton events weird, our search strategy can then capture most of the signal. The scalar widths are small enough so that the narrow-width approximation holds. Consequently, we proceed as follows. We vary $m_k$ and $m_h$ in the range 100, 200, ..., 1000 GeV. For each pair, and having fixed all couplings to the values mentioned before, we compute Monte Carlo events and estimate the efficiency ($\epsilon$) for selecting events in each of the categories described in previous sections. The number of expected signal events in each category for a given luminosity ($\mathcal{L}$) can then be computed as
\begin{equation}
 N = \sigma(pp\rightarrow k^{++} k^{--})\times \mathcal{L} \times\epsilon~.
\end{equation}
Then, we compare again the three most sensitive categories (one for each SR) with the corresponding SM background. The results are shown in figure~\ref{fig:nh}.
The grey triangle at the bottom of the plot is already excluded by neutrino and low-energy data.
The green region enclosed by the solid green line is the area that can be excluded using the already collected luminosity $\mathcal{L} = 70$ fb$^{-1}$ (counting 35 for each experiment, ATLAS and CMS). It is worth noting that, if we take $\kappa > 5$, the triangle excluded by neutrino data goes down to 400 GeV. Consequently, current LHC data can already constrain regions not bounded before by other experiments.
Likewise, the orange region enclosed by the dashed orange line is the region that can be excluded in a high-luminosity phase of the LHC with $\mathcal{L} = 3000$ fb$^{-1}$. Doubly-charged scalar masses as large as 1 TeV could be probed for $k^{\pm\pm}\rightarrow \ell^\pm \ell^\pm$, but also in exotic decays. 

\subsubsection{Inverted hierarchy}
The IH parameter space of the Zee-Babu model is very constrained by neutrino data, being $m_k$ and $m_h$ out of the LHC reach for most values of $\phi$ and $\delta$. These parameters stand for the physical Majorana	and Dirac phases in the PMNS matrix, respectively. (Note that, in the Zee-Babu model, one of the neutrinos is massless.) However, these bounds are significantly weakened if $\phi\sim\delta\sim\pi$, and even smaller for large values of $\kappa$. For definiteness, we take
\begin{align}
 g_{11} &= g_{23} =  0.1, g_{12} = g_{22} = g_{13} = g_{33} = 0.0001, \\
 f_{12} &= 0.1, f_{13} = -0.1, f_{23} = 0.01, \kappa = 5. 
\end{align}
These values are allowed by current data, even for small values of $m_k, m_h\sim 100$ GeV~\cite{Herrero-Garcia:2014hfa}.
We emphasize that there is very small room for variations in this hierarchy of couplings. Moreover, although $\mathcal{O}(1)$ modifications in their absolute values might be in principle allowed, the expected scalar branching ratios, and therefore the LHC phenomenology, would remain the same.
\begin{figure}[t]
\begin{center}
  \includegraphics[width=0.95\columnwidth]{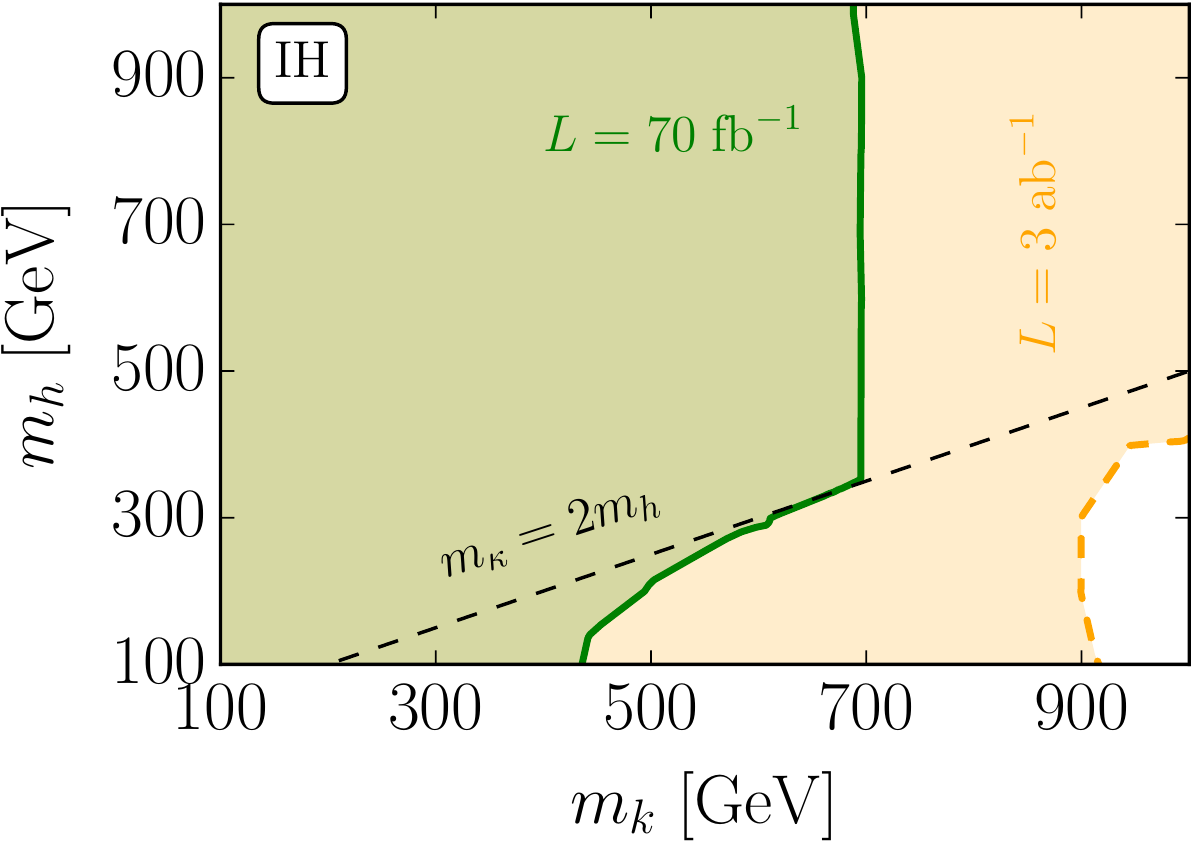}
 \end{center}
 \caption{\it Excluded regions in the plane $m_k-m_h$ in the IH in the Zee-Babu model. See the text for details.}\label{fig:ih}
\end{figure}
We proceed as in the NH case and test which region of the $m_k-m_h$ plane can be probe with current and future LHC data. The result is shown in figure~\ref{fig:ih}. The smaller region for $m_k < 2m_h$ in comparison with the NH case is due to the smaller $k$ branching ratio into leptons. Again, there are non-previously bounded regions that can be excluded with the current LHC data, even for $m_k > 2m_h$. Likewise, masses in the TeV region could be tested with future analyses.

\section{Conclusions}\label{sec:conclusions}
We have argued that current LHC analyses can not probe doubly-charged scalars with exotic decays, as those arising in models of radiatively induced neutrino masses. Novel searches, as the ones proposed in this article, combining different signal regions and observables, are however sensitive to these particles. Thus, masses as large as $\gtrsim 500$ GeV can be accessed with current LHC data for doubly-charged scalars $k$ decaying as $k^{\pm\pm}\rightarrow h^\pm h^\pm$, with $h^{\pm}\rightarrow \ell^\pm \nu$. These numbers are only slightly smaller than those for doubly-charged scalars decaying into pairs of same-sign leptons. This result has important implications for concrete scenarios, most importantly the Zee-Babu model. In particular, we have shown that parameter space regions of this model not yet constrained by neutrino and low-energy experiments can be tested with current LHC data, while much larger regions could be excluded at the 95 \% C.L. in a high-luminosity phase.

Conversely, models in which $k$ decays predominately via the emission of $W$ bosons are by far less constrained.%

In any case, our results (most importantly the selection of signal regions and observables, as well as the precise determination of the SM background) can be applied to very different models of neutrino masses. Therefore, we expect that this work will be of interest for many forthcoming studies.
\vspace{-0.2cm}
\section*{ Acknowledgments}
\noindent
This work is partially supported by the Spanish MINECO under grant FPA2014-54459-P, by the Severo Ochoa Excellence Program under grant SEV-2014-0398.
and by the “Generalitat Valenciana” under grant GVPROMETEOII2014-087. MC thanks the MITP for hospitality during the completion of this work.

%
\vspace{-0.2cm}
\appendix
\section{Tables}\label{sec:app1}
The number of background  for an integrated luminosity of $\mathcal{L} = 13.9$ fb$^{-1}$ in the different categories of SR1, SR2 and SR3 are shown in tables~\ref{tab:11}, \ref{tab:12} and \ref{tab:13}, tables~\ref{tab:21}, \ref{tab:22} and \ref{tab:23} and tables~\ref{tab:31}, ~\ref{tab:32} and ~\ref{tab:33}, respectively.
\begin{table*}[ht!]
 \begin{center}
\begin{adjustbox}{width=0.5\textwidth}
\begin{tabular}{||c|c|c|c|c|c|c|c|c|c||}\hline
              & & &  & & & & & & \\[-0.2cm]
     $m_{\ell\ell}\diagdown S_T > $ [GeV]         &  100 & 200 & 300 & 400 & 500 & 600 & 700 & 800 & 900  \\
              &                               & & & & & & & & \\[-0.2cm]
 \hline
 & & &   & & & & & & \\[-0.2cm]
 100 &  13000 & 2000 & 570 & 180 & 76 & 34   & 16   & 7   & 4.2 \\
 200 &   2700 & 1600 & 530 & 180 & 72 & 33   & 15   & 7   & 4.2 \\
 300 &    900 &  700 & 470 & 170 & 70 & 32   & 14   & 6.7 & 4.1 \\
 400 &    370 &  320 & 250 & 150 & 65 & 31   & 14   & 6.7 & 4.1 \\
 500 &    180 &  160 & 130 &  96 & 61 & 29   & 13   & 6.7 & 4   \\
 600 &     89 &   84 &  72 &  58 & 41 & 27   & 13   & 6.5 & 4   \\
 700 &     46 &   45 &  40 &  34 & 24 & 17   & 12   & 6.3 & 4   \\
 800 &      26 &   26 &  25 &  20 & 16 & 12   &  9   & 5.6 & 3.9 \\
 900 &      16 &   16 &  15 &  14 & 11 &  7.8 &  6.5 & 5.2 & 3.7 \\
 \hline
 \end{tabular}
 \end{adjustbox}
 \end{center}
 \caption{\it $S_T$ versus $m_{\ell\ell}$ in SR1. %
 }\label{tab:11}
\end{table*}
\begin{table*}[ht!]
 \begin{center}
\begin{adjustbox}{width=0.5\textwidth}
\footnotesize
\begin{tabular}{||c|c|c|c|c|c|c|c|c|c||}\hline
              & & &  & & & & & & \\[-0.2cm]
     $m_T\diagdown S_T > $ [GeV]         &  100 & 200 & 300 & 400 & 500 & 600 & 700 & 800 & 900 \\
              &                              & & & & & & & & \\[-0.2cm]
 \hline
 & & &   & & & & & & \\[-0.2cm]
 100 &  14000 & 2000 & 580 & 190 & 76 & 34 & 16   & 7   & 4.2 \\
 200 &   5100 & 1800 & 550 & 180 & 74 & 33 & 15   & 7   & 4.2 \\
 300 &   1600 & 1100 & 520 & 170 & 72 & 32 & 15   & 6.7 & 4.1 \\
 400 &    630 &  510 & 340 & 170 & 69 & 32 & 14   & 6.7 & 4.1 \\
 500 &    290 &  250 & 200 & 120 & 66 & 30 & 14   & 6.7 & 4.1 \\
 600 &    140 &  130 & 110 &  76 & 51 & 29 & 13   & 6.7 & 4.1 \\
 700 &      73 &   70 &  61 &  48 & 32 & 21 & 13   & 6.7 & 4   \\
 800 &        42 &   41 &  38 &  32 & 24 & 16 & 11   & 6.4 & 4   \\
 900 &        25 &   25 &  24 &  22 & 17 & 12 &  8.7 & 6   & 3.9 \\
 \hline
 \end{tabular}
 \end{adjustbox}
 \end{center}
 \caption{\it $S_T$ versus $m_T$ in SR1.
 The cut on $m_T$ applies to the two reconstructed transverse masses.
 }\label{tab:12}
\end{table*}
\begin{table*}[ht!]
 \begin{center}
\begin{adjustbox}{width=0.5\textwidth}
\footnotesize
\begin{tabular}{||c|c|c|c|c|c|c|c|c|c||}\hline
              & & &  & & & &  & &\\[-0.2cm]
     $m_{T2}\diagdown S_T > $ [GeV]        &  100 & 200 & 300 & 400 & 500 & 600 & 700 & 800 & 900  \\
              &                              & & & & & & & & \\[-0.2cm]
 \hline
 & &  & & & & & & & \\[-0.2cm]
 100 &       19     &    5.5   &   1.7   &   0.65  &  0.2   &  0.11  &  0.063 & 0.056 & 0.047 \\
 200 &         1.1   &    0.97  &   0.55  &   0.21  &  0.08  &  0.051 &  0.036 & 0.031 & 0.02  \\
 300 &         0.19  &    0.19  &   0.17  &   0.058 &  0.047 &  0.03  &  0.02  & 0.02  & 0.017 \\
 400 &         0.035 &    0.035 &   0.034 &   0.034 &  0.024 &  0.021 &  0.022 & 0.02  & 0.017 \\
 500 &         0.021 &    0.021 &   0.021 &   0.021 &  0.021 &  0.018 &  0.017 & 0.017 & 0.017 \\
 600 &         0.018 &    0.018 &   0.018 &   0.018 &  0.018 &  0.018 &  0.017 & 0.017 & 0.017 \\
 700 &         0.017 &    0.017 &   0.017 &   0.017 &  0.017 &  0.017 &  0.017 & 0.017 & 0.017 \\
 800 &         0.017 &    0.017 &   0.017 &   0.017 &  0.017 &  0.017 &  0.017 & 0.017 & 0.017 \\
 900 &         0     &    0     &   0     &   0     &  0     &  0     &  0     & 0     & 0     \\
 \hline
 \end{tabular}
 \end{adjustbox}
 \end{center}
 \caption{\it $S_T$ versus $m_{T2}$ in SR1.
 }\label{tab:13}
\end{table*}

%
\begin{table*}[ht]
 \begin{center}
\begin{adjustbox}{width=0.5\textwidth}
\footnotesize
\begin{tabular}{||c|c|c|c|c|c|c|c|c|c||}\hline
              & &  & & & & & & & \\[-0.2cm]
     $m_{\ell\ell}\diagdown S_T > $ [GeV]        &  100 & 200 & 300 & 400 & 500 & 600 & 700 & 800 & 900  \\
              &                              & & & & & & & & \\[-0.2cm]
 \hline
 & &   & & & & & & & \\[-0.2cm]
 100 &  2200   & 740   & 240   &  92   & 42   & 21   & 11    & 6.3  & 3.9 \\
 200 &   570   & 350   & 160   &  71   & 35   & 19   &  9.7  & 6    & 3.8 \\
 300 &   180   & 130   &  85   &  47   & 26   & 15   &  8.4  & 5.3  & 3.7 \\
 400 &    67   &  57   &  43   &  30   & 19   & 12   &  7.4  & 4.7  & 3.3 \\
 500 &    30   &  28   &  22   &  17   & 12   &  8.5 &  5.2  & 3.7  & 2.9 \\
 600 &    14   &  14   &  11   &   9.6 &  7.3 &  5.6 &  3.7  & 2.8  & 2.2 \\
 700 &     6.8 &   6.8 &   6.1 &   5.4 &  4.3 &  3.6 &  2.6  & 2.2  & 1.8 \\
 800 &     3.6 &   3.6 &   3.5 &   3.3 &  2.7 &  2.4 &  1.9  & 1.8  & 1.4 \\
 900 &     2.2 &   2.2 &   2.2 &   2   &  1.5 &  1.4 &  0.98 & 0.91 & 0.9 \\
 \hline
 \end{tabular}
 \end{adjustbox}
 \end{center}
 \caption{\it $S_T$ versus $m_{\ell\ell}$ in SR2. %
 }\label{tab:21}
\end{table*}
\begin{table*}[t]
 \begin{center}
\begin{adjustbox}{width=0.5\textwidth}
\footnotesize
\begin{tabular}{||c|c|c|c|c|c|c|c|c|c||}\hline
              & &  & & & & & & & \\[-0.2cm]
     $m_T\diagdown S_T > $ [GeV]       &  100 & 200 & 300 & 400 & 500 & 600 & 700 & 800 & 900 \\
              &                               & & & & & & & & \\[-0.2cm]
 \hline
 & & &   & & & & & & \\[-0.2cm]
 100 &   840     & 370     & 140     &  61     & 30     & 16     &  8.2   & 5.1  & 3.5   \\
 200 &    100     &  92     &  54     &  30     & 17     & 10     &  5.8   & 3.9  & 2.7   \\
 300 &     22     &  22     &  19     &  13     &  8.5   &  5.7   &  3.6   & 2.6  & 1.6   \\
 400 &      6.7   &   6.7   &   6.6   &   5.8   &  4.1   &  2.8   &  1.9   & 1.4  & 0.96  \\
 500 &      2.3   &   2.3   &   2.3   &   2.3   &  1.9   &  1.6   &  1.2   & 1    & 0.64  \\
 600 &     0.99  &   0.99  &   0.99  &   0.97  &  0.9   &  0.74  &  0.57  & 0.54 & 0.45  \\
 700 &      0.42  &   0.42  &   0.42  &   0.42  &  0.42  &  0.41  &  0.39  & 0.39 & 0.32  \\
 800 &      0.12  &   0.12  &   0.12  &   0.12  &  0.12  &  0.12  &  0.11  & 0.11 & 0.1   \\
 900 &      0.049 &   0.049 &   0.049 &   0.049 &  0.049 &  0.049 &  0.049 & 0.05 & 0.045 \\
 \hline
 \end{tabular}
 \end{adjustbox}
 \end{center}
 \caption{\it $S_T$ versus $m_T$ in SR2.
 The cut on $m_T$ applies to the two reconstructed transverse masses.
 }\label{tab:22}
\end{table*}
\begin{table*}[t]
 \begin{center}
\begin{adjustbox}{width=0.5\textwidth}
\footnotesize
\begin{tabular}{||c|c|c|c|c|c|c|c|c|c||}\hline
              & &  & & & & &  & &\\[-0.2cm]
     $m_{T2}\diagdown S_T > $ [GeV]        &  100 & 200 & 300 & 400 & 500 & 600 & 700 & 800 & 900  \\
              &                               & & & & & & & & \\[-0.2cm]
 \hline
 & & &  & & & & & &  \\[-0.2cm]
 100 &  2400   & 800   & 250   &  95   & 43   & 22   & 11   & 6.4  & 4    \\
 200 &   600   & 370   & 170   &  76   & 37   & 19   & 10   & 6.2  & 3.9  \\
 300 &   180   & 140   &  91   &  51   & 28   & 16   &  8.9 & 5.5  & 3.7  \\
 400 &    69   &  59   &  44   &  31   & 20   & 13   &  7.7 & 4.8  & 3.5  \\
 500 &    31   &  28   &  23   &  17   & 13   &  8.8 &  5.4 & 3.9  & 3    \\
 600 &    15   &  14   &  11   &   9.8 &  7.5 &  5.8 &  3.9 & 3    & 2.3  \\
 700 &     6.9 &   6.9 &   6.2 &   5.5 &  4.4 &  3.7 &  2.7 & 2.3  & 1.9  \\
 800 &     3.6 &   3.6 &   3.6 &   3.4 &  2.7 &  2.5 &  1.9 & 1.8  & 1.5  \\
 900 &     2.2 &   2.2 &   2.2 &   2   &  1.5 &  1.4 &  1   & 0.94 & 0.93 \\
 \hline
 \end{tabular}
 \end{adjustbox}
 \end{center}
 \caption{\it $S_T$ versus $m_{T2}$ in SR2.
 }\label{tab:23}
\end{table*}
%
\vspace{15cm}
\pagebreak
\begin{table*}[ht]
 \begin{center}
\begin{adjustbox}{width=0.5\textwidth}
\footnotesize
\begin{tabular}{||c|c|c|c|c|c|c|c|c|c||}\hline
              & &  & & & & & & & \\[-0.2cm]
     $m_{\ell\ell}\diagdown S_T > $ [GeV]         &  100 & 200 & 300 & 400 & 500 & 600 & 700 & 800 & 900  \\
              &              &                  & & & & & & & \\[-0.2cm]
 \hline
  & &  & & & & & & & \\[-0.2cm]
 100 &    58     &  44     & 21     & 10     & 5.2   & 2.9   & 1.5   & 1     & 0.54  \\
 200 &     6.8   &   6     &  5.2   &  4.2   & 3.2   & 2.1   & 1.2   & 0.79  & 0.5   \\
 300 &     1.5   &   1.5   &  1.4   &  1.2   & 1.1   & 0.85  & 0.64  & 0.49  & 0.31  \\
 400 &     0.48  &   0.48  &  0.44  &  0.41  & 0.35  & 0.3   & 0.27  & 0.24  & 0.15  \\
 500 &     0.15  &   0.15  &  0.15  &  0.15  & 0.15  & 0.13  & 0.13  & 0.13  & 0.073 \\
 600 &     0.053 &   0.053 &  0.053 &  0.053 & 0.052 & 0.069 & 0.069 & 0.069 & 0.035 \\
 700 &     0.052 &   0.052 &  0.052 &  0.052 & 0.051 & 0.051 & 0.051 & 0.051 & 0.017 \\
 800 &     0.017 &   0.017 &  0.017 &  0.017 & 0.017 & 0.017 & 0.017 & 0.017 & 0     \\
 900 &     0.017 &   0.017 &  0.017 &  0.017 & 0.017 & 0.017 & 0.017 & 0.017 & 0     \\
 \hline
 \end{tabular}
 \end{adjustbox}
 \end{center}
 \caption{\it $S_T$ versus $m_{\ell\ell}$ in SR3. %
 }\label{tab:31}
\end{table*}
\begin{table*}[ht]
 \begin{center}
\begin{adjustbox}{width=0.5\textwidth}
\footnotesize
\begin{tabular}{||c|c|c|c|c|c|c|c|c|c||}\hline
              & & & & & & &  & & \\[-0.2cm]
     $m_T\diagdown S_T > $ [GeV]      &   100 & 200 & 300 & 400 & 500 & 600 & 700 & 800 & 900 \\
              &                               & & & & & & & & \\[-0.2cm]
 \hline
 & & &  & & & & &  & \\[-0.2cm]
 100 &    65     &  48     & 23     & 11     & 5.4   & 3     & 1.6   & 1     & 0.57   \\
 200 &     9.4   &   8.5   &  6.8   &  5.2   & 3.6   & 2.3   & 1.3   & 0.87  & 0.52   \\
 300 &     2.4   &   2.3   &  2.1   &  1.8   & 1.5   & 1.1   & 0.76  & 0.58  & 0.34   \\
 400 &     0.78  &   0.78  &  0.75  &  0.69  & 0.55  & 0.42  & 0.37  & 0.32  & 0.18   \\
 500 &    0.22  &   0.22  &  0.22  &  0.23  & 0.22  & 0.2   & 0.17  & 0.16  & 0.083  \\
 600 &     0.1   &   0.1   &  0.1   &  0.1   & 0.094 & 0.12  & 0.11  & 0.11  & 0.047  \\
 700 &    0.069 &   0.069 &  0.069 &  0.069 & 0.068 & 0.066 & 0.063 & 0.059 & 0.013  \\
 800 &    0.037 &   0.037 &  0.037 &  0.037 & 0.037 & 0.037 & 0.037 & 0.034 & 0.0047 \\
 900 &     0.02  &   0.02  &  0.02  &  0.02  & 0.02  & 0.02  & 0.02  & 0.02  & 0.0024 \\
 \hline
 \end{tabular}
 \end{adjustbox}
 \end{center}
 \caption{\it $S_T$ versus $m_T$ in SR3.
 The cut on $m_T$ applies to the two reconstructed transverse masses.
 }\label{tab:32}
\end{table*}
\begin{table*}[ht]
 \begin{center}
\begin{adjustbox}{width=0.5\textwidth}
\footnotesize
\begin{tabular}{||c|c|c|c|c|c|c|c|c|c||}\hline
              & & & &  & & &  & &\\[-0.2cm]
     $m_{T2}\diagdown S_T > $ [GeV]         &  100 & 200 & 300 & 400 & 500 & 600 & 700 & 800 & 900  \\
              &                               & & & & & & & & \\[-0.2cm]
 \hline
 & & &   & & & & & & \\[-0.2cm]
 100 &  220    & 120    & 34    & 13    & 6.3  & 3.2  & 1.7  & 1.1  & 0.61 \\
 200 &  56    &  45    & 26    & 12    & 6.1  & 3.2  & 1.7  & 1.1  & 0.61 \\
 300 &  18    &  16    & 12    &  8.3  & 5.2  & 3    & 1.6  & 1.1  & 0.61 \\
 400 &    6.7  &   6.2  &  5    &  4.2  & 3.3  & 2.3  & 1.5  & 1.1  & 0.62 \\
 500 &    3.2  &   3.1  &  2.7  &  2.4  & 2.1  & 1.6  & 1.2  & 0.95 & 0.59 \\
 600 &     1.6  &   1.6  &  1.5  &  1.4  & 1.2  & 0.96 & 0.87 & 0.73 & 0.48 \\
 700 &    0.88 &   0.88 &  0.88 &  0.82 & 0.72 & 0.59 & 0.56 & 0.52 & 0.36 \\
 800 &     0.6  &   0.6  &  0.6  &  0.58 & 0.5  & 0.43 & 0.41 & 0.39 & 0.28 \\
 900 &     0.46 &   0.46 &  0.46 &  0.43 & 0.37 & 0.33 & 0.31 & 0.29 & 0.2  \\
 \hline
 \end{tabular}
 \end{adjustbox}
 \end{center}
 \caption{\it $S_T$ versus $m_{T2}$ in SR3.
 }\label{tab:33}
\end{table*}

\clearpage
\bibliographystyle{JHEP}
\bibliography{notes-new}{}

\end{document}